\documentclass[twocolumn,showpacs,showkeys,preprintnumbers]{revtex4}
\usepackage{amssymb}
\usepackage{amsfonts}
\usepackage{amsmath}
\usepackage{graphicx}
\usepackage{dcolumn}
\usepackage{bm}
\usepackage{epsfig}

\begin{document}

\title{Polytropic representation of non-isotropic kinetic pressure tensor\\
for non-ideal\ plasma fluids in relativistic jets}
\author{Claudio Cremaschini}
\affiliation{Research Centre for Theoretical Physics and Astrophysics, Institute of
Physics, Silesian University in Opava, Bezru\v{c}ovo n\'{a}m.13, CZ-74601
Opava, Czech Republic}
\date{\today }

\begin{abstract}
Non-ideal fluids are likely to be affected by the occurrence of pressure
anisotropy effects, whose understanding for relativistic systems requires
knowledge of the energy-momentum tensor. In this paper the case of
magnetized jet plasmas at equilibrium is considered, in which both
microscopic velocities of constituent particles as well as the continuum
fluid flow are treated as relativistic ones. A theoretical framework based
on covariant statistical kinetic approach is implemented, which permits the
proper treatment of single-particle and phase-space kinetic constraints and,
ultimately, the calculation of the system continuum fluid fields associated
with physical observables. A Gaussian-like solution for the kinetic
distribution function (KDF) is constructed, in which the physical mechanism
responsible for the generation of temperature anisotropy is identified with
magnetic moment conservation. A Chapman-Enskog representation of the same
KDF is then obtained in terms of expansion around an equilibrium isotropic
Juttner distribution. This permits the analytical calculation of the fluid $%
4-$flow and stress-energy tensor and the consequent proof that the
corresponding kinetic pressure tensor is non-isotropic. As a notable result,
the validity of a polytropic representation for the perturbative
non-isotropic pressure contributions is established, whereby directional
pressures exhibit specific power-law functional dependences on fluid density.
\end{abstract}

\keywords{Non-ideal magnetized fluids; Relativistic plasma jets; Equation of
state;\ \ Pressure anisotropy; Polytropic representation.\\
\underline{Corresponding Author:} Claudio Cremaschini - email:
claudiocremaschini@gmail.com}
\maketitle

\section{Introduction}

A relevant theoretical issue in statistical physics concerns the
mathematical description of the complex phenomenology that characterizes
microscopic as well as continuum fluid dynamics of magnetized plasmas at
equilibrium and which exhibit properly-defined symmetric geometric
structures. Applications include in particular astrophysical scenarios
represented by relativistic jets associated with compact objects, binary
compact object mergers and gamma-ray bursts \cite%
{Sironi,colpi1,colpi2,colpi3} as well as the physics of complex non-linear
magnetized plasmas in the framework of soft-matter studies \cite%
{Gourdain,Byvank,Mond,Salafia}. For these systems, both single-particle and
macroscopic fluid velocities of the plasma can become relativistic, i.e.,
comparable with the speed of light $c$ when measured in a defined reference
frame (e.g., the laboratory or fluid co-moving frames). In addition, they
can be subject to the simultaneous presence of electromagnetic (EM) and
possible gravitational fields, determined respectively by the Maxwell
equations and by the Einstein equations of General Relativity. These
circumstances require therefore adoption of relativistic covariant
theoretical approaches to be carried out in either flat or curved space-time
according to the specific physical connotations of each scenario \cite%
{Hamlin}.

In this regard, two different types of statistical descriptions can be
envisaged for the study of plasma systems composed by a number $N\gg 1$ of
species charged particles. The first one is represented by relativistic
microscopic and kinetic theories, which are rooted on single-particle
dynamics and conservation laws and yield information about the statistical
distribution of matter in phase-space $\Omega =\mathcal{R}^{4}\times 
\mathcal{V}^{4}$, where $\mathcal{R}^{4}$ and $\mathcal{V}^{4}$ denote
respectively the unconstrained four-dimensional particle configuration and
velocity spaces. The fundamental quantity of kinetic theories is represented
by the species kinetic distribution function (KDF) $f_{\mathrm{s}}\equiv f_{%
\mathrm{s}}(\mathbf{x},s)$, which is defined in phase-space $\Omega $ and is
subject to a precise kinetic evolution equation that is characteristic of
the plasma regime and the type of phenomena to be studied. Here, $s$ and $%
\mathbf{x}\equiv \left( r^{\mu },u^{\mu }\equiv dr^{\mu }/ds\right) $ denote
respectively the single-particle proper time and state, namely the $4-$%
position and the velocity $4-$vector, while the subscript "$s$" stands for
the species index (e.g., electrons and ions). The second approach is
provided by fluid theories, e.g., relativistic hydrodynamic (HD) or
magnetohydrodynamic (MHD) treatments, whereby the plasma is described as a
continuum in the configuration space $\mathcal{R}^{4}$ and is characterized
by observable fluid fields (e.g., mass and energy density, temperature and
velocity fields) that satisfy a precise set of fluid equations, such as the
continuity, Euler-momentum and Fourier equations \cite{Maha01,Maha02}.
However, fluid theories suffer a problem of completeness, namely they
require necessarily an independent prescription of a suitable finite set of
closure conditions warranting the integrability of fluid equations. Examples
of closure conditions include the equation of state for the fluid pressure
or pressure-tensor, the heat-flux vector and the viscosity tensor. On the
other hand, when kinetic theories are available, fluid models can always be
determined "a posteriori" from them, by means of statistical average
integration of the KDF and kinetic equation over velocity space $\mathcal{V}%
^{4}$. Remarkably, the knowledge of the KDF can also resolve at the same
time the closure-condition problem, so that the prescription of the closure
conditions for the fluid equations becomes unique and self-consistent in
this framework. Accordingly, the kinetic approach provides the most complete
statistical information about microscopic and macroscopic plasma dynamics.

A crucial feature of plasma dynamics that calls for the adoption of a
kinetic formulation originates from the fact that the latter systems are
likely to develop phase-space anisotropies, which would be otherwise
scarcely understandable by a fluid treatment ignoring the underlying kinetic
properties. The issue is pertinent mainly to collisionless regimes, namely
species plasma states existing on time-scales shorter than the
characteristic species collision time scales defined in a given reference
system. Alternatively, collisionless plasmas can be identified by the
requirement that the species mean free path of charged particles, denoted
with $\lambda _{mfp}$, is much greater than the largest characteristic scale
length of the plasma $L$, to be defined for example as in Eq.(\ref{L-scale})
below. Thus, collisionless fluids are such that $\lambda _{mfp}\gg L$. We
refer instead to the weakly-collisional fluid regime if the inequality $%
\lambda _{mfp}\gtrsim L$ applies. In practice, the collisionless assumption
can be verified a posteriori once the kinetic solution is known by explicit
evaluation of the species scale-length $\lambda _{mfp}$. The latter must
take into consideration the rate of occurrence of microscopic collisions
among fluid particles, possibly in combination with the simultaneous action
of confining mechanisms associated with magnetic fields (e.g., the Larmor
rotation).

In the collisionless case, the fundamental dynamical equation evolving the
KDF is represented by the covariant Vlasov equation, which in Lagrangian
form is written as%
\begin{equation}
\frac{d}{ds}f_{\mathrm{s}}\left( \mathbf{x}\left( s\right) ,s\right) =0,
\end{equation}%
where\ in general the function $f_{\mathrm{s}}$\ can still depend explicitly
on $s$ and $\mathbf{x}\left( s\right) $\ is the single-particle phase-space
trajectory parametrized in terms of the particle proper-time $s$. However,
as explained below, phase-space anisotropies should not be excluded also in
collisional plasmas, where they might still occur as perturbative
contributions to the isotropic Maxwellian KDF that customarily identifies
these systems. In fact, several physical mechanisms can be identified to be
responsible for the onset of phase-space anisotropies. These include for
example relativistic single-particle dynamics and the constraints placed by
related conservation laws, like in the case of gyrokinetic theory predicting
occurrence of adiabatic invariants, particle confinement mechanisms, such as
Larmor rotation or epicyclic motion, plasma kinetic regimes, boundary-layer
effects as well as presence of viscosity mechanisms, radiation fields or
multi-species mutual interactions \cite{Ott,Mika,Kunz,Egedal,Soni}.

The statistical treatment of these features is essential in order to
characterize the dynamical and thermodynamical properties of the systems of
interest. For example, phase-space anisotropies of the type mentioned above
can be responsible for the occurrence of collective drift-velocities in
strongly-magnetized plasmas, to be associated with diamagnetic, finite
Larmor-radius or energy-conservation effects. In addition, phase-space
anisotropies can generate also shear-flow phenomena characterized by strong
gradients of angular velocity in axisymmetric systems, magnetic field
generation (kinetic dynamo)\ by local current densities in plasmas, as well
as peculiar statistical configurations referred to as non-symmetric or
energy-independent kinetic equilibria. Similarly, temperature and pressure
anisotropies as well as non-vanishing heat fluxes can arise on this basis 
\cite{Tavecchio}.

The issue about the form that the pressure tensor can assume and the
physical meaning of its content are of particular interest for the present
study. In this connection, different physical sources of velocity-space
anisotropies for the KDF have been pointed out in plasmas, which determine
correspondingly non-isotropic pressure tensors. The most relevant one is
enhanced by the adiabatic conservation of the guiding-center particle
magnetic moment predicted by gyrokinetic theory, which is associated with
the Larmor rotation of charged particles in magnetic fields. The same type
of physical mechanism is expected to operate also for relativistic plasmas,
as suggested by the relativistic kinetic theory established in recent
contributions. This concerned respectively the proof of existence of kinetic
equilibria of relativistic collisionless plasmas in the presence of
non-stationary electromagnetic fields and the covariant formulation of
spatially non-symmetric kinetic equilibria in magnetized plasmas \cite%
{pop1,pop2}. Additional phase-space anisotropies responsible for the
generation of temperature anisotropy have been also identified with the
conservation of particle canonical momentum, diamagnetic and
energy-correction effects in non-uniform plasmas (see for example Refs.\cite%
{Loop,APJ}) and also with peculiar space-time kinematic constraints due to
conservation of Killing tensors affecting neutral matter in curved
space-time, like in the case of the Kerr metric \cite{Carter,Tolman}.

In light of these developments, it is correct to state that in plasma
physics and in fluid dynamics in general, in the absence of an independent
evolution equation for the fluid scalar pressure $P$, the prescription of
the equation of state (EoS) or either the pressure tensor appears generally
a complex problem of crucial importance. This involves representing the same
fluid pressure in terms of independent fluid fields (i.e., the fluid state)
and external parameters describing the physical, dynamical and
thermodynamical properties of the fluid system under consideration \cite%
{Kaur}. The problem becomes even more serious in the case of relativistic
plasma regimes, due to the existence of additional kinematic constraints for
particle dynamics, boost effects due to definition of coordinate systems as
well as the representation and integration of the KDF, which altogether
reflect themselves on the practical calculation of the EoS. Most frequently,
the EoS is expressed in terms of a single isotropic scalar pressure $P$
prescribed by an equation of the type $P=P\left( n,T,\mathbf{V}\right) $.
This may depend both on the local fluid number density $n$, the local scalar
temperature $T$ and also the local fluid velocity field $\mathbf{V}$ defined
in the fluid co-moving frame. Its precise form should in principle be
determined separately based on phenomenological models and/or microscopic
(i.e., kinetic) physics that pertains the structure and interactions
occurring among the same constituents of the fluid, possibly subject to
external fields \cite{Eyal,Poor,Muir}. A particular case that belongs to
this category pertains to ideal fluids \cite{Li}, to be intended as
continuum systems described at microscopic level by a phase-space statistics
determined by a local and possibly Maxwellian KDF. For ideal systems of this
type the pressure $P\ $is a position-dependent scalar function expressed by
the well-known ideal relationship (in IS dimensional units)$\ P=nT$.

An alternative model route often pursued in hydrodynamics consists in
treating the scalar pressure $P$ as a function of the fluid mass density $%
\rho $ only \cite{Tejeda1,Rhodes,Caprioli1,Ovalle}. This yields the
so-called polytropic form of the EoS, which in customary notation is written
as%
\begin{equation}
P=\kappa \rho ^{\Gamma },  \label{poly-1}
\end{equation}%
where $\kappa $ is a suitable dimensional numerical factor of
proportionality (frequently taken to be constant), while $\Gamma $ is the
polytropic exponent factor (or polytropic index). The polytropic
representation can be applicable to either non-relativistic and relativistic
systems, including gases, plasmas and also degenerate matter if justified on
physical basis and microscopic dynamics \cite{Boccelli,coll-2016,Liva}.

However, actual physical fluid systems may be expected to deviate in several
ways from the ideal-fluid state. This can be particularly relevant in the
case of collisionless or suitably weakly-collisional astrophysical charged
fluids and/or magnetized plasmas \cite{Liva2}. In fact, in these systems,
the action of gravitational and electromagnetic fields in combination with
other effects like radiation fields, conservation laws, dissipation or
trapping phenomena, boundary-layer conditions and geometrical
configuration-space constraints can lead to the onset of so-called non-ideal
fluids \cite{Asenjo1,Zhu,F2023}. In the present framework a fluid is said to
be non-ideal if its EoS ceases to be represented by a scalar pressure and
becomes instead expressed in terms of a pressure tensor exhibiting
directional pressures, and therefore yielding a pressure (and temperature)
anisotropy \cite{Hall,Holm,Douanla}. Hence, the non-ideal case amounts to
replacing the pressure $P$ with a momentum or stress-energy tensor of the
type%
\begin{equation}
P\rightarrow \Pi _{\mu \nu },  \label{pressure tensor}
\end{equation}%
whose spatial section, when evaluated in the fluid co-moving frame,
identifies the corresponding non-isotropic pressure tensor characterized by
distinctive directional pressure components. The proper understanding of the
physical effects contributing to the generation of a pressure tensor and the
correct determination of its mathematical structure demand the adoption of
statistical treatments in terms of appropriate kinetic theories. In fact,
from the kinetic point of view, the occurrence of a pressure tensor (\ref%
{pressure tensor})\ that cannot be reduced to a scalar pressure component
corresponds to a continuum fluid described by a non-isotropic, namely
non-Maxwellian, KDF. Accordingly, this type of occurrence identifies a
non-ideal fluid. In such a framework, deviations from a Maxwellian function
must be treated in statistical way as they generally amount to the onset of
phase-space anisotropies that ultimately show up as macroscopic properties
characteristic of non-ideal fluids.

\subsection{Goals of the paper}

Given these premises, in this paper the characterization of pressure
anisotropy effect and related prescription of the EoS for relativistic
magnetized plasmas at equilibrium forming symmetric jet structures is
considered,\ which correspond to non-ideal fluids. Specifically, the goals
of the present research can be summarized according to the following scheme:

1) The definition of the theoretical framework in terms of microscopic
single-particle dynamics, covariant statistical kinetic theory and the
mathematical relationship with the corresponding continuum fluid
description. This includes, in particular, the statistical formulation of
isotropic and non-isotropic kinetic equilibria and the development of
covariant gyrokinetic theory permitting the representation of particle
magnetic moment.

2)\ The analytical construction of an exact equilibrium solution for the
KDF, obtained by implementing the method of invariants. This amounts to
representing the KDF in terms of a Gaussian-like solution which is
prescribed in such a way to depend (explicitly and implicitly) only on the
integrals of motion and adiabatic invariants available for the relativistic
plasma. This permits the selection of the physical mechanism responsible for
the generation of temperature anisotropy in the relativistic jet plasma of
interest, which is identified here with magnetic moment conservation for
charged particles in the prescribed magnetic field.

3) The perturbative treatment of the relativistic Gaussian-like equilibrium
KDF by means of a suitable Chapman-Enskog expansion carried out around a
leading-order equilibrium isotropic distribution, denoted as Juttner or
relativistic Maxwellian KDF. The outcome provides an analytical series
representation of the full equilibrium solution in which the non-isotropic
contributions are singled-out as first-order polynomial terms weighted over
the Juttner Gaussian function.

4) The explicit calculation of the stress-energy tensor for the relativistic
plasma, evaluated in the fluid co-moving frame. Two physical applications
are then considered. The first one concerns the evaluation of the energy
density of the non-ideal plasma fluid, which is provided by the time-time
component of the same stress-energy tensor in the co-moving frame. It is
shown that the energy density is reduced with respect to the purely
isotropic configuration, as a consequence of the statistical phase-space
anisotropy. The second one concerns instead the representation of the
pressure tensor, or equivalently of the fluid tensorial equation of state,
which is identified with the space-space components of the stress-energy
tensor in the co-moving frame. In this regard, the mathematical proof of the
non-isotropic character of the pressure tensor is given.

5)\ The representation of the pressure tensor and, specifically, of the
perturbative non-isotropic pressure contributions in polytropic form. This
is reached thanks to a suitable iterative scheme\ holding in accordance to
the Chapman-Enskog series and starting from a relation of the type (\ref%
{poly-1}), whereby both the scalar pressure and mass density are those
associated with the Juttner KDF.

The research contains analytical work concerning equilibrium solutions for
relativistic non-ideal plasma fluids. The solution proposed is obtained
under validity of suitable assumptions for the plasma state, the system
geometric features and structure of electromagnetic fields. It is understood
that the present analytical solution can certainly provide a convenient
framework for more involved future studies that might take into account
additional details of the complex plasma phenomenology. These considerations
apply both to equilibrium and dynamically-evolving scenarios, where the
model precision as well as the model applicability should be verified
according to each specific physical realization. In this regard, concerning
the implementation of numerical techniques, we mention the so-called
structure-preserving method which can provide an effective way to perform
such a verification, see for example Refs.\cite{Ch1,Ch2,Ch3,Ch4}.

Concerning the notations, we follow Ref.\cite{degroot}, with the difference
that throughout the next calculations we set for the particle rest-mass $%
m_{o}=1$ and speed of light $c=1$. This permits to have an identity between
velocity and momentum spaces, implying in turn simpler expressions. This
choice nevertheless does not affect the physical content of the solution. In
fact, one can always restore the correct dimensional units of physical
observables by restoring "a posteriori" the correct powers of $m_{o}$ and $c$
based on dimensional analysis.

\section{Non-isotropic kinetic and fluid solutions}

In this section we introduce the concepts of isotropic and non-isotropic
kinetic and fluid solutions for collisionless $N-$body systems in the
framework of the covariant Vlasov theory. These refer both to the properties
of the KDF as well as to the form of the corresponding fluid stress-energy
tensor $T^{\mu \nu }\left( r^{\alpha }\right) $. In order to address the
issue and without loss of generality, the subscript $\mathrm{j}$ of the
species index will be omitted in this section.

The generic particle has a $4-$position $r^{\mu }$ and a $4-$velocity $%
u^{\mu }\equiv \frac{dr^{\mu }}{ds}$, with $s$ being the corresponding
world-line proper time. The line element $ds$ satisfies the identity%
\begin{equation}
ds^{2}=g_{\mu \nu }\left( r^{\alpha }\right) dr^{\mu }dr^{\nu },  \label{ds}
\end{equation}%
where $g_{\mu \nu }\left( r^{\alpha }\right) \in \mathcal{R}^{4}$ is the
position-dependent metric tensor which is taken to have signature $g_{\mu
\nu }=\left( +,-,-,-\right) $, while $\mathcal{R}^{4}$ is the $4-$%
dimensional smooth Lorentzian manifold. Eq.(\ref{ds}) implies the mass-shell
constraint for the $4-$velocity $u^{\mu }$, namely%
\begin{equation}
g_{\mu \nu }\left( r^{\alpha }\right) u^{\mu }u^{\nu }=1.
\label{mass-shell1}
\end{equation}%
This must be intended as identically satisfied along the particle geodesic
trajectory determined by the equation of motion%
\begin{equation}
\frac{D}{Ds}u^{\mu }\left( s\right) =0,
\end{equation}%
where $\frac{D}{Ds}$ denotes the covariant derivative and the proper-time
parametrization is adopted for\ $u^{\mu }$, consistent with Eq.(\ref{ds}).
We then denote the particle Lagrangian state $\mathbf{x}\left( s\right) $
with $\mathbf{x}\left( s\right) \equiv \left( r^{\mu }\left( s\right)
,u^{\mu }\left( s\right) \equiv \frac{dr^{\mu }\left( s\right) }{ds}\right) $%
.

For the covariant representation of the particle dynamics and the subsequent
treatment of kinetic equilibria we adopt a formalism based on the
introduction of a tetrad of unit $4-$vectors. Thus, we introduce the set of
orthogonal unit $4-$vectors $\left( a^{\mu },b^{\mu },c^{\mu },d^{\mu
}\right) $ which are defined by the coordinate system, where $a^{\mu }$ and $%
\left( b^{\mu },c^{\mu },d^{\mu }\right) $ are respectively time-like and
space-like. In terms of the unit $4-$vectors representation the particle $4-$%
velocity can therefore be decomposed as%
\begin{equation}
u^{\mu }\equiv u_{0}a^{\mu }+u_{1}b^{\mu }+u_{2}c^{\mu }+u_{3}d^{\mu },
\label{4-veeldeco}
\end{equation}%
to be denoted as tetrad representation, where $\left(
u_{0},u_{1},u_{2},u_{3}\right) $ are the corresponding components and are
separately $4-$scalars. Then, we recall that for single-particle dynamics
the $4-$dimensional velocity space $\mathcal{V}^{4}$ is defined as the
tangent bundle of the configuration-space manifold $\mathcal{R}^{4}$ whose $%
4-$vectors are subject to the mass-shell constraint (\ref{mass-shell1}).
Thus, the unconstrained 8-dimensional phase-space $\Omega $ is defined as $%
\Omega =\mathcal{R}^{4}\times \mathcal{V}^{4}$. By construction, the tetrad
representation provides a local mutual relationship among the components of
the $4-$velocity (\ref{4-veeldeco}). In particular, this yields the
representation%
\begin{equation}
u_{0}=\sqrt{1+u_{1}^{2}+u_{2}^{2}+u_{3}^{2}},  \label{u-zero}
\end{equation}%
so that the time-component of the $4-$velocity depends on the squared
space-components. Notice that this equation is related to the mass-shell
condition (\ref{mass-shell1}).

Based on these premises, we can proceed with the appropriate definitions of
isotropic and non-isotropic kinetic solutions. We consider first the
distribution function. In the following a generic relativistic KDF $f$ will
be said to be isotropic on the velocity space $\mathcal{V}^{4}$ if it
carries even powers of all the $4-$velocity components $\left(
u_{1},u_{2},u_{3}\right) $ and this functional dependence is isotropic. A
particular case of isotropic dependence is through the velocity component $%
u_{0}$, since the relationship (\ref{u-zero}) always applies locally. Hence,
a KDF of the type $f=f\left( u_{0}\right) $ represents an isotropic
distribution. A notable example of isotropic KDF of this form is the
relativistic Maxwellian (or Juttner)\ function $f_{J}$. An exhaustive
treatment of the Juttner solution can be found in Ref.\cite{degroot}. In
contrast, a KDF $f$ will be said to be non-isotropic on $\mathcal{V}^{4}$ if
it exhibits a non-isotropic dependence on the even powers of the $4-$%
velocity components $\left( u_{1},u_{2},u_{3}\right) $. Examples of
non-isotropic KDF of this type for relativistic plasma kinetic equilibria
can be found in Refs.\cite{pop1,pop2}. It is clear from these considerations
that non-isotropic distribution functions necessarily differ from local
Maxwellian distributions, while from a statistical point of view such
deviations are due to velocity-space, or more generally phase-space kinetic
effects arising from single-particle dynamics.

As mentioned above, in kinetic theory the observable fluid fields of a given 
$N-$body system can be determined \textquotedblleft a
posteriori\textquotedblright\ once the kinetic solution for the KDF and its
dynamical behavior are known. This is accomplished by defining the same
fluid fields as appropriate velocity integrals of the KDF weighted by
suitable phase-space dependent weight functions. This approach marks the
strict connection between kinetic and fluid approaches and gives physical
meaning to the corresponding mathematical solutions. We consider the
stress-energy tensor $T^{\mu \nu }\left( r^{\alpha }\right) $ for the
continuum fluid system, to be defined by the $4-$velocity integral%
\begin{equation}
T^{\mu \nu }\left( r^{\alpha }\right) =2\int_{\mathcal{V}^{4}}\sqrt{-g}%
d^{4}u\Theta \left( u^{0}\right) \delta \left( u^{\mu }u_{\mu }-1\right)
u^{\mu }u^{\nu }f,  \label{tmunu}
\end{equation}%
where the Dirac-delta takes into account the kinematic constraint (\ref%
{mass-shell1}) for the $4-$velocity when performing the integration, $\Theta 
$ denotes the Theta-function selecting the root of $u^{0}$, while $\sqrt{-g}$
is the square-root of the determinant of the background metric tensor.
Invoking the tetrad representation for the $4-$velocity (\ref{4-veeldeco}),
the integral (\ref{tmunu}) can be reduced to%
\begin{equation}
T^{\mu \nu }\left( r^{\alpha }\right) =\int_{\mathcal{V}^{3}}\frac{\sqrt{-g}%
d^{3}u}{\sqrt{1+u_{1}^{2}+u_{2}^{2}+u_{3}^{2}}}u^{\mu }u^{\nu }f,
\label{tmunu-bis}
\end{equation}%
which is now defined over the $3-$dimensional tangent space in which the
component $u_{0}$ of the $4-$velocity must be intended as dependent on the
other components according to Eq.(\ref{u-zero}). In the following, a
stress-energy tensor $T^{\mu \nu }\left( r^{\alpha }\right) $ is said to be
isotropic if it is defined in terms of an isotropic KDF $f$.
Correspondingly, the same tensor field will be referred to as non-isotropic
stress-energy tensor if the KDF corresponding to the collisionless $N-$body
system and entering Eq.(\ref{tmunu-bis}) is a non-isotropic distribution, in
the sense defined above. Thus, an example of isotropic stress-energy tensor
is the one associated with a relativistic Maxwellian (or Juttner) KDF $f_{J}$
which describes ideal fluids, to be denoted as $T_{J}^{\mu \nu }\left(
r^{\alpha }\right) $. This takes the customary tensorial form%
\begin{equation}
T_{J}^{\mu \nu }\left( r^{\alpha }\right) =n_{J}e_{J}U^{\mu }U^{\nu
}-P\Delta ^{\mu \nu },  \label{T-max}
\end{equation}%
where $n_{J}$ is the particle number density and $e_{J}$ is the energy per
particle carried by the same Juttner KDF, so that $ne$ is the fluid energy
density, $U^{\mu }$ is the fluid $4-$velocity, $P$ is the scalar isotropic
pressure and $\Delta ^{\mu \nu }$ is the so-called projector operator $%
\Delta ^{\mu \nu }=g^{\mu \nu }-U^{\mu }U^{\nu }$.

\section{\textbf{Covariant gyrokinetic theory}}

From a statistical point of view, the concept of temperature anisotropy
refers to the occurrence of an anisotropy in the directional particle
velocity dispersions averaged over the KDF and measured with respect to a
given reference system. An analogous definition applies to pressure
anisotropy. In collisionless astrophysical plasmas different sources of
temperature anisotropy have been pointed out at equilibrium, which combine
the contributions arising from single-particle dynamics and non-uniform
background gravitational and EM fields. The most relevant one is a
magnetic-related effect induced by the conservation of particle magnetic
moment as predicted by gyrokinetic (GK) theory. The reason is that this
mechanism relies on the velocity-space symmetry of the Larmor rotation
motion, and therefore it is independent of the existence of additional
space-time symmetries. In the following we recall fundamental results of
covariant GK theory which are needed for the present research, while a
detailed treatment of its variational formulation can be found in Refs.\cite%
{pop2,Bek1,Bek2} for relativistic particle dynamics and in Ref.\cite{Cr2013}
for the corresponding non-relativistic limit.

The non-perturbative formulation of covariant GK theory is achieved in terms
of a guiding-center transformation of particle state of the form%
\begin{eqnarray}
r^{\mu } &=&r^{\prime \mu }+\rho _{1}^{\prime \mu },  \label{1} \\
u^{\mu } &=&u^{\prime \mu }\oplus \nu _{1}^{\prime \mu },  \label{2}
\end{eqnarray}%
where $r^{\prime \mu }$ is the guiding-center position $4-$vector, $\rho
_{1}^{\prime \mu }$ is referred to as the relativistic Larmor $4-$vector, $%
\oplus $ denotes the relativistic composition law which warrants that $%
u^{\mu }$ is a $4-$velocity and primed quantities are all evaluated at $%
r^{\prime \mu }$. A background EM field described by the antisymmetric
Faraday tensor $F_{\mu \nu }$ is assumed. Denoting with $H$\ and $E$\ the $%
4- $scalar eigenvalues of $F_{\mu \nu }$, the latter can be conveniently
represented as%
\begin{equation}
F_{\mu \nu }=H\left( c_{\nu }d_{\mu }-c_{\mu }d_{\nu }\right) +E\left(
b_{\mu }a_{\nu }-b_{\nu }a_{\mu }\right) .  \label{tetrad-fmunu}
\end{equation}%
which determines the orientation of the tetrad basis (EM-tetrad frame). The
physical meaning is that $H$\ and $E$\ coincide with the observable magnetic
and electric field strengths in the reference frame where the electric and
the magnetic fields are parallel. We shall assume that locally the inequality%
\begin{equation}
E^{2}-H^{2}<0  \label{E-H}
\end{equation}%
is always satisfied. Notice furthermore that the EM-tetrad basis represents
the natural covariant generalization of the magnetic-related triad system
formed by the orthogonal right-handed unit 3-vectors $\left( \mathbf{e}_{1},%
\mathbf{e}_{2},\mathbf{e}_{3}\equiv \mathbf{b}\right) $\textbf{\ }usually
introduced in non-relativistic treatments, where $\mathbf{b}$ is the unit
vector parallel to the local direction of the magnetic field and $\left( 
\mathbf{e}_{1},\mathbf{e}_{2}\right) $ are two unit orthogonal vectors in
the normal plane to $\mathbf{b}$. When the $4-$vector $u^{\prime \mu }$ is
projected on the guiding-center EM-basis, it determines the representation%
\begin{equation}
u^{\prime \mu }\equiv u_{0}^{\prime }a^{\prime \mu }+u_{\parallel }^{\prime
}b^{\prime \mu }+w^{\prime }\left[ c^{\prime \mu }\cos \phi ^{\prime
}+d^{\prime \mu }\sin \phi ^{\prime }\right] ,  \label{u-primo}
\end{equation}%
which defines the gyrophase angle $\phi ^{\prime }$ associated with the
Larmor rotation, where $w^{\prime }$ is the magnitude of $u^{\prime \mu }$
in the plane $\pi _{\perp }\equiv (c^{\prime \mu },d^{\prime \mu })$. Then,
denoting $\left\langle {}\right\rangle _{\phi ^{\prime }}\equiv \frac{1}{%
2\pi }\oint d\phi ^{\prime }$ the gyrophase-averaging operator, the
following non-perturbative representation is obtained for the relativistic
particle magnetic moment $m^{\prime }$:%
\begin{equation}
m^{\prime }=\left\langle \frac{\partial \rho _{1}^{\prime \mu }}{\partial
\phi ^{\prime }}\left[ \left( u_{\mu }^{\prime }\oplus \nu _{1\mu }^{\prime
}\right) +qA_{\mu }\right] \right\rangle _{\phi ^{\prime }},  \label{m-exact}
\end{equation}%
where $q\equiv \frac{Ze}{mc^{2}}$, $Ze$ is the particle electric charge and $%
A_{\mu }\left( r\right) $ is the EM $4-$potential, with $r$ denoting the
generic configuration-space coordinate belonging to the domain occupied by
the plasma.

Based on the exact GK solution, a covariant perturbative theory can be
developed for the analytical asymptotic treatment of Eq.(\ref{m-exact}),
yielding a perturbative representation of $m^{\prime }$. This is achieved by
a Larmor-radius expansion of the relevant dynamical quantities in terms of
the $4-$scalar dimensionless parameter%
\begin{equation}
\varepsilon \equiv \frac{r_{L}^{\prime }}{L},  \label{epsilon}
\end{equation}%
to be assumed infinitesimal, namely $\varepsilon \ll 1$. Here $r_{L}^{\prime
}\equiv \sqrt{\rho _{1}^{\prime \mu }\rho _{1\mu }^{\prime }}$, while $L$
denotes a characteristic invariant scale-length associated with the
background fields. More precisely, if $\Lambda \equiv \Lambda \left(
r\right) $ identifies an arbitrary suitably-defined scalar fluid field,
which is constructed in terms of the tensor fluid fields characterizing the
system (for example, the fluid density or temperature), the Lorentz scalar $%
L $ can be conveniently defined as%
\begin{equation}
\frac{1}{L^{2}}\equiv \sup \left[ \frac{1}{\Lambda ^{2}}\partial _{\mu
}\Lambda \partial ^{\mu }\Lambda \right] .  \label{L-scale}
\end{equation}%
Here, the $\sup $ is performed with respect to $\Lambda \left( r\right) $,
while $\partial _{\mu }$ denotes the tensor partial derivative operator.
Notice that in principle the parameter $\varepsilon $ is species-dependent,
although we omit the species subscript "s" in the following to avoid heavy
notations and without possible misunderstandings. The simultaneous validity
of the two inequalities $\varepsilon \ll 1$ and (\ref{E-H}) identifies the
local condition of \emph{strongly-magnetized relativistic plasma}, whereby
its physical properties change on scale-lengths greater than the Larmor
radius. In fact, if the EM Lorentz scalar $\mathbf{E}\cdot \mathbf{H}$
vanishes locally, there is a Lorentz frame where $\mathbf{E}=0$ so that only
the magnetic field $\mathbf{H}$ is present. In the following, however, we
shall allow for greater generality $\mathbf{E}\cdot \mathbf{H}\sim O\left(
\varepsilon ^{0}\right) $. We notice here that, for a strongly-magnetized
plasma, an alternative equivalent definition of the parameter $\varepsilon $
can also be given. In fact, one can assume $\varepsilon \equiv \frac{r_{L}}{L%
}$, where $r_{L}\equiv \sqrt{\rho _{1}^{\mu }\rho _{1\mu }}$ and $\rho
_{1}^{\mu }$ is the Larmor $4-$vector evaluated at particle position rather
than at guiding-center. In such a case $\rho _{1}^{\mu }$ and $\rho
_{1}^{\prime \mu }$ differ by infinitesimal contributions, so that, given
validity of the present ordering assumption, one can write $\rho _{1}^{\mu
}=\rho _{1}^{\prime \mu }+O\left( \varepsilon \right) $.

Upon carrying out the perturbative expansion to first order in $\varepsilon $%
, one obtains that in the perturbative theory the $4-$vector $u^{\prime \mu
} $ in Eq.(\ref{u-primo}) represents the leading-order particle $4-$%
velocity. Similarly, the asymptotic representation%
\begin{equation}
m^{\prime }=\mu ^{\prime }\left[ 1+O\left( \varepsilon \right) \right]
\label{m'}
\end{equation}%
is reached, where%
\begin{equation}
\mu ^{\prime }=\frac{w^{\prime 2}}{2qH^{\prime }}  \label{mu'}
\end{equation}%
is the leading-order contribution of the magnetic moment which represents an
adiabatic invariant, with $H^{\prime }$ denoting the corresponding
guiding-center eigenvalue of $H$. From the point of view of the asymptotic
formulation of GK theory, thanks to the asymptotic conservation of $\mu
^{\prime }$, particle dynamics in the plane $\pi _{\perp }\equiv (c^{\prime
\mu },d^{\prime \mu })$ orthogonal to the local direction of magnetic field
decouples from that in the plane $\pi _{\parallel }\equiv (a^{\prime \mu
},b^{\prime \mu })$ containing the parallel space direction to it. This
feature is recovered also in the non-relativistic regime of GK dynamics
between the direction $\mathbf{b}$ and the orthogonal plane $\left( \mathbf{e%
}_{1},\mathbf{e}_{2}\right) $. Finally, we notice that, since the
leading-order expression of $m^{\prime }$ depends only on $w^{\prime 2}$ and
not on the remaining component $u_{\parallel }^{\prime }$, it follows that,
according to the definitions introduced above, $\mu ^{\prime }$ can rightly
represent a source of phase-space anisotropy for the kinetic equilibrium,
producing a non-isotropic KDF. This generates a corresponding non-isotropic
stress-energy tensor and the statistical phenomenon of temperature
anisotropy defined with respect to the planes $\pi _{\perp }$ and $\pi
_{\parallel }$.

\section{Symmetries, conservation laws and adiabatic invariants}

In the present treatment we must distinguish between two particular inertial
reference frames, respectively identified with the laboratory and the fluid
co-moving frames. In both inertial frames a coordinate system $\left(
ct,R,\varphi ,z\right) $ is introduced, with $t$ being the coordinate time
and $\left( R,\varphi ,z\right) $ denoting spatial cylindrical coordinates.
The two frames are related by a Lorentz transformation in terms of the
constant plasma fluid $4-$velocity $U^{\mu }$. The latter is subject to the
kinematic constraint%
\begin{equation}
U^{\mu }U_{\mu }=1.  \label{4vel1}
\end{equation}%
In the co-moving frame the $4-$velocity $U^{\mu }$ takes the form $U_{\text{%
com}}^{\mu }=\left( 1,0,0,0\right) $. Instead, in the inertial laboratory
frame it is assumed that the same $4-$velocity is represented as%
\begin{equation}
U^{\mu }=\left( U^{t},0,0,U^{z}\right) ,
\end{equation}%
so that, when seen by an observer, the plasma possesses spatial flow
velocity directed along the $z$-axis. Notice that the consistency of such a
representation must be checked \textquotedblleft a
posteriori\textquotedblright\ in terms of the kinetic solution to be
determined below. Hence, Eq.(\ref{4vel1}) implies the relationship $%
U^{t2}=1+U^{z2}$ holding in the laboratory frame. Concerning the plasma
components, the generic particle has a $4-$velocity $u^{\mu }\equiv \frac{%
dr^{\mu }}{ds}$ and related $4-$momentum $p^{\mu }\equiv m_{o}u^{\mu }$,
with $m_{o}$ denoting the particle rest-mass. From the kinematic constraint (%
\ref{mass-shell1}) it follows analogously that $p_{\mu }p^{\mu }=m_{o}^{2}$.
We then denote by $P^{\mu }=(u^{\mu }+qA^{\mu })$ the particle generalized $%
4-$momentum, where $A^{\mu }=\left( \Phi ,\mathbf{A}\right) $ is the EM $4-$%
potential, so that the $4-$scalar $P_{\mu }U_{\text{com}}^{\mu }=P_{t}$
identifies the particle energy in the co-moving frame. We recall that for
convenience of calculation we set hereon $m_{o}=1$, implying identity
between velocity and momentum spaces.

The representation of the EM $4-$potential $A^{\mu }$ demands the
introduction of a suitable asymptotic ordering. In detail, we shall assume
that the EM $4-$potential is an analytic function of $\varepsilon $ which
can be represented as a Laurent series of the form%
\begin{equation}
A^{\mu }=\frac{1}{\varepsilon }A_{-1}^{\mu }+\varepsilon ^{0}A_{0}^{\mu
}+O\left( \varepsilon \right) ,  \label{laurent}
\end{equation}%
where the lower index labels the order-$O\left( \varepsilon ^{i}\right) $ of
the series expansion for $i=-1,0,1,..$. The total EM $4-$potential $A^{\mu }$
is then decomposed as%
\begin{equation}
A^{\mu }=A^{\left( ext\right) \mu }+A^{\left( pl\right) \mu }+A^{\left(
RR\right) \mu },  \label{A-total}
\end{equation}%
where $A^{\left( ext\right) \mu }$ is the external field generated in the
plasma domain by external sources, $A^{\left( pl\right) \mu }$ yields the
self-generated plasma EM field due to collective plasma currents, while $%
A^{\left( RR\right) \mu }$ stands for the single-particle EM
radiation-reaction (EM-RR) component. It is important to stress in fact
that, contrary to the case of non-relativistic systems, for relativistic
plasmas it is generally not possible to exclude the contribution of
single-particle EM radiation emission occurring through EM-RR dynamics \cite%
{EPJ2,EPJ5,Brodin}. Nevertheless, its contribution must be properly set in
order to permit existence of kinetic equilibria. Thus, in validity of Eq.(%
\ref{laurent}), different relative orderings among the three EM $4-$%
potentials can be assumed for the corresponding leading-order contributions
entering the expansion (\ref{laurent}). For example, if the dominant term in
(\ref{laurent}) is produced by the\ local equilibrium collective plasma
currents one can set%
\begin{eqnarray}
A^{\left( ext\right) \mu } &\sim &O\left( \varepsilon ^{k}\right) ,k\geq 0,
\label{ord1} \\
A^{\left( pl\right) \mu } &\sim &O\left( \frac{1}{\varepsilon }\right) , \\
A^{\left( RR\right) \mu } &\sim &O\left( \varepsilon ^{k_{s}}\right)
,k_{s}>k,  \label{ord3}
\end{eqnarray}%
where $k$ and $k_{s}$ are two real numbers. Instead, when the external EM
field is the dominant component a possible ordering can be%
\begin{eqnarray}
A^{\left( ext\right) \mu } &\sim &O\left( \frac{1}{\varepsilon }\right) ,
\label{e1} \\
A^{\left( pl\right) \mu } &\sim &O\left( \varepsilon ^{k}\right) ,k\geq 0, \\
A^{\left( RR\right) \mu } &\sim &O\left( \varepsilon ^{k_{s}}\right)
,k_{s}>k.  \label{e3}
\end{eqnarray}%
In both cases, however, the contribution due to EM-RR is considered to be of
highest order in the relative ordering scheme. This condition is physically
motivated. In fact, from one side EM-RR cannot be ruled out in relativistic
plasmas, but at the same time this must represent a perturbative correction
in order to warrant existence of kinetic equilibria, to be intended
necessarily in asymptotic sense.

A crucial issue to establish at this point concerns the proof that, as a
matter of principle, exact conservation laws are necessarily excluded in the
presence of EM-RR. As a result, only asymptotic conservation laws can
possibly exist, depending on the ordering assumption imposed on $A^{\left(
RR\right) \mu }$. Details about the mathematical proof are reported for
completeness in the Appendix. The immediate consequence of the conclusion
reached in the Appendix is that for relativistic collisionless plasmas only
adiabatic invariants can exist. In the remaining part of this section we
discuss the case of invariants associated with configuration-space
symmetries, to be complemented by the investigation of velocity-space
invariants obtained by gyrokinetic theory in the previous section. More
precisely, we assume that in the inertial reference frames considered here
the coordinates $t$ and $z$ identify symmetry (i.e.,
asymptotically-ignorable) coordinates for the plasma state and EM fields.
Let us denote the external and collective EM $4-$potentials accordingly as $%
A^{\left( ext\right) \mu }=A^{\left( ext\right) \mu }\left( \varepsilon
^{k}ct,R,\varphi ,\varepsilon ^{k}z\right) $ and $A^{\left( pl\right) \mu
}=A^{\left( pl\right) \mu }\left( \varepsilon ^{k}ct,R,\varphi ,\varepsilon
^{k}z\right) $, where for convenience of notation we have retained a unique
parameter $k$ in setting the ordering dependences on ignorable coordinates.
A generalization to different indices is immediate. It follows that the
canonical momenta associated with the time coordinate $t$ and the spatial
coordinate $z$, namely $P_{t}$ and $P_{z}$, are adiabatic invariants of $%
O\left( \varepsilon ^{k}\right) $ defined by%
\begin{eqnarray}
P_{t} &\equiv &m_{o}c\frac{dr_{t}}{ds}+\frac{q}{c}\left( \overline{A}%
_{t}^{(ext)}+\overline{A}_{t}^{(pl)}+2\overline{A}_{t}^{(RR)}(r)\right) , \\
P_{z} &\equiv &m_{o}c\frac{dr_{z}}{ds}+\frac{q}{c}\left( \overline{A}%
_{z}^{(ext)}+\overline{A}_{z}^{(pl)}+2\overline{A}_{z}^{(RR)}(r)\right) .
\end{eqnarray}%
In terms of the components of the $4-$vector potential and the particle $4-$%
velocity detailed above, $P_{t}$ and $P_{z}$ become%
\begin{eqnarray}
P_{t} &\equiv &m_{o}c\gamma +\frac{q}{c}\Phi +2\frac{q}{c}\overline{A}%
_{t}^{(RR)}(r),  \label{Pt} \\
P_{z} &\equiv &m_{o}\gamma v_{z}+\frac{q}{c}A_{z}+2\frac{q}{c}\overline{A}%
_{z}^{(RR)}(r).  \label{Pz}
\end{eqnarray}%
For definiteness, in the following we shall require that the slow-dependence
on the $z$-coordinate entering the external $4-$vector components $\overline{%
A}_{t}^{(ext)}$ and $\overline{A}_{z}^{(ext)}$ is assumed to be intrinsic in
the nature of the external field. Instead, the slow-dependences on the time
coordinate and the $z$-coordinate entering the collective-plasma $4-$vector
components $\overline{A}_{t}^{(pl)}$ and $\overline{A}_{z}^{(pl)}$ are
assumed to be due only to the RR effect associated with $\overline{A}_{\mu
}^{(RR)}$.

In view of these considerations and invoking the Laurent relative orderings
given above it is possible to represent $P_{t}$ and $P_{z}$ also as follows%
\begin{eqnarray}
P_{t} &=&P_{t}^{\left( 0\right) }+\varepsilon ^{k}P_{t}^{\left( 1\right) },
\label{pt0} \\
P_{z} &=&P_{z}^{\left( 0\right) }+\varepsilon ^{k}P_{z}^{\left( 1\right) },
\label{pz0}
\end{eqnarray}%
where $P_{t}^{\left( 0\right) }\equiv m_{o}c\gamma +\frac{q}{c}\Phi $, $%
P_{t}^{\left( 1\right) }\equiv 2\frac{q}{c}\overline{A}_{t}^{(RR)}(r)$,
while $P_{z}^{\left( 0\right) }\equiv m_{o}\gamma v_{z}+\frac{q}{c}\psi _{T}$
and $P_{z}^{\left( 1\right) }\equiv 2\frac{q}{c}\overline{A}_{z}^{(RR)}(r)$
respectively. In this decomposition the two terms $P_{t}^{\left( 0\right) }$
and $P_{z}^{\left( 0\right) }$ are the leading-order contributions which
exhibit asymptotic time and $z$ symmetries, while the perturbative terms $%
P_{t}^{\left( 1\right) }$ and $P_{z}^{\left( 1\right) }$ are due to the
EM-RR effect and contribute to determine the intrinsic slow $t-$ and $z-$%
variations of the system. This feature allows us to construct kinetic
equilibria which are slowly-varying in space and time as a consequence of
the RR phenomenon, which can be investigated as a separate issue in future
studies.

\section{Relativistic kinetic equilibrium}

A qualitative feature of magnetized plasmas, in particular astrophysical jet
plasmas, is related to the occurrence of kinetic plasma regimes which
persist for long times (with respect to the observer and/or plasma
characteristic times), despite the presence of macroscopic time-varying
phenomena of various origin, such as flows, non-uniform gravitational/EM
fields and EM radiation, possibly including that arising from
single-particle radiation-reaction processes. It is reasonable to argue that
these states might actually correspond - at least locally and in a suitable
asymptotic sense - to some kind of exact or asymptotic stationary
configuration, so that the plasma KDF may correspond to so-called kinetic
equilibria.\ This means that, when referred to a relativistic reference
frame, all the species KDFs $f_{\mathrm{s}}$\ are required to be smooth,
strictly-positive ordinary functions which depend only on a suitable set of
invariants $I_{\ast }(\mathbf{x})$, namely such that $\frac{d}{ds}I_{\ast }(%
\mathbf{x}(s))=0+O\left( \varepsilon ^{k}\right) $. Here, $\mathbf{x}\left(
s\right) $\ is the single-particle phase-space trajectory parametrized in
terms of the particle proper-time $s$, while $\varepsilon \ll 1$\ is the
characteristic small dimensionless parameter defined in (\ref{epsilon}).
More precisely, following the discussion reported above, for $k=0$ the real
scalar functions $I_{\ast }(\mathbf{x})$\ identify exact integrals of
motion, i.e., invariant phase functions which do not explicitly depend on $s$%
, while for $k\geq 1$\ they identify adiabatic invariants of order $k$,
namely asymptotically-conserved phase functions that vary in proper-time on
time-scales much longer than the characteristic time-scales of the system.
According to the conclusions reached in Section IV, for relativistic plasmas
only adiabatic invariants have a physical meaning as representatives of
concrete physical systems. Therefore, in this sense kinetic equilibria may
arise also in physical scenarios in which macroscopic fluid fields (e.g.,
fluid $4-$flows) and/or the EM field might be time dependent when seen from
an observer reference frame. It is worth noting that in practice this may
require only that the corresponding KDF actually changes slowly in time
(with respect to an appropriate microscopic time-scale), so that from the
phenomenological viewpoint these states can still be regarded as
quasi-stationary or even non-stationary.

In this section we develop the formalism for the representation of
analytical solutions of the KDF corresponding to kinetic equilibria
describing non-ideal fluid configurations. As shown below, these types of
kinetic solutions are found to differ from the purely-Maxwellian solution.
Given these premises, we can now consider the problem of the construction of
kinetic equilibria for collisionless relativistic jet plasmas in the
framework of Vlasov-Maxwell theory. These are expressed by the corresponding
species equilibrium KDFs, to be denoted with the symbol $f_{\ast \mathrm{s}}$%
. By definition the equilibrium species KDF $f_{\ast \mathrm{s}}$ must be an
asymptotic solution of the Vlasov equation, namely%
\begin{equation}
r_{L}\frac{d}{ds}\ln f_{\ast \mathrm{s}}\left( \mathbf{x}(s)\right)
=0+O\left( \varepsilon ^{k+1}\right) ,
\end{equation}%
so that it is an adiabatic invariant of $O\left( \varepsilon ^{k+1}\right) $%
. The previous condition can be satisfied identically by the requirement
that $f_{\ast \mathrm{s}}$ must be assumed function only of the relevant
adiabatic invariants, and therefore to be defined in a suitable inertial
Lorentz frame. This technique is known in plasma physics as the method of
invariants. From the results of Sections III and IV we have that, for the
case of interest here, the set of invariants is represented by the magnetic
moment $m^{\prime }$ due to velocity-space symmetry law and the components ($%
P_{t},P_{z}$) of particle $4-$momentum arising from space-time configuration
symmetries, namely%
\begin{equation}
I_{\ast }(\mathbf{x}(s))=\left( P_{t},P_{z},m^{\prime }\right) .
\end{equation}%
It is understood that the components of particle $4-$momentum must be
specified according to the reference frame. Therefore, the general
representation for the functional dependence of the equilibrium KDF $f_{\ast 
\mathrm{s}}$ is necessarily of the type%
\begin{equation}
f_{\ast \mathrm{s}}=f_{\ast \mathrm{s}}\left( I_{\ast }(\mathbf{x}%
(s))\right) =f_{\ast \mathrm{s}}\left( P_{t},P_{z},m^{\prime }\right) .
\label{form-2}
\end{equation}%
We notice that the dependence on particle energy $P_{t}$ generates the
typical Gaussian (i.e., Maxwellian) character of the solution, while that on
the linear momentum $P_{z}$ permits inclusion of spatial longitudinal
velocity\ $U^{z}$. Instead, the dependence on $m^{\prime }$ is ultimately
associated with the occurrence of pressure anisotropy. It expresses the
non-ideal character of the solution and its deviation from the
purely-Maxwellian case.

In the following we are interested in looking for the possible existence of
Gaussian-like relativistic equilibria for jet plasmas. A realization of this
type includes in particular the case of Maxwellian-like solutions for the
KDF. Since the plasma is assumed to be in collisionless regime, it is always
possible to represent the equilibrium KDF as a sum of $N$\ sub-species KDFs\ 
$f_{\ast \mathrm{s}}^{\left( j\right) }$, with $j=1,N$, so that $f_{\ast 
\mathrm{s}}=\sum_{j}f_{\ast \mathrm{s}}^{\left( j\right) }$. In this way,
for each $f_{\ast \mathrm{s}}^{\left( j\right) }$\ the kinetic approach in
terms of method of invariants applies, allowing for their representation as
relativistic Gaussian-like distributions. This kind of formalism is
necessary and appropriate to treat relativistic plasma jets and gamma ray
bursts, which are likely to be characterized by the co-existence of
different sub-populations of collisionless plasmas.

In particular, adopting such sub-species decomposition for $f_{\ast \mathrm{s%
}}$, the target is reached by requiring that each $f_{\ast \mathrm{s}%
}^{\left( j\right) }$ is assumed to be asymptotically close (in a sense to
be later defined) to a local isotropic relativistic Maxwellian KDF $f_{J%
\mathrm{s}}^{\left( j\right) }$. The latter is characterized by non-uniform
fluid fields, identified with isotropic temperature $T_{\mathrm{s}}^{\left(
j\right) }$, number density $n_{\mathrm{s}}^{\left( j\right) }$ and $4-$%
velocity $U^{\mu \left( j\right) }$ of the form $U^{\mu \left( j\right)
}=\left( U^{t\left( j\right) },0,0,U^{z\left( j\right) }\right) $ in the
laboratory frame. For each $f_{\ast \mathrm{s}}^{\left( j\right) }$ the
fluid $4-$velocity $U^{\mu \left( j\right) }$ is assumed to be a local
adiabatic invariant, namely to depend at most slowly on configuration-space
coordinates as $U^{\mu \left( j\right) }=U^{\mu \left( j\right) }\left(
\varepsilon ^{k}r\right) $, with $k\geq 1$. Finally, the KDF $f_{\ast 
\mathrm{s}}$ must be a strictly-positive real function and it must be
summable, in the sense that the tensorial velocity moments on the
velocity-space $\mathcal{V}^{4}$ of the general form%
\begin{equation}
\Xi _{\mathrm{s}}\left( r^{\alpha }\right) =\int_{\mathcal{V}^{4}}\sqrt{-g}%
d^{4}u\Theta \left( u^{0}\right) \delta \left( u^{\mu }u_{\mu }-1\right) K_{%
\mathrm{s}}(u^{\mu })f_{\ast \mathrm{s}},  \label{velomo}
\end{equation}%
must exist for a suitable ensemble of weight functions $\left\{ K_{\mathrm{s}%
}(u^{\mu })\right\} $, to be prescribed in terms of polynomials of arbitrary
degree defined with respect to the particle velocity $4-$vector. This can be
warranted by imposing the minimum requirement that the equilibrium KDF must
have a Gaussian-like dependence on particle momentum component $P_{t}$.

Given these prescriptions, a particular solution for the equilibrium KDF is
taken of the form%
\begin{equation}
f_{\ast \mathrm{s}}=N_{\ast s}\exp \left[ -\gamma _{\ast s}P_{\mu }U^{\mu
}-\alpha _{\ast s}m^{\prime }\right] ,  \label{f*}
\end{equation}%
which we refer to as the \emph{equilibrium relativistic non-isotropic KDF}.
Notice that hereafter the sub-species index "$j$" will be omitted for
clarity of notation and without possibility of misunderstandings. Here, the $%
4-$vector $U^{\mu }$ is by assumption a local adiabatic invariant, while the 
$4-$scalars%
\begin{equation}
\Lambda _{\ast s}\equiv \left( N_{\ast s},\gamma _{\ast s},\alpha _{\ast
s}\right)
\end{equation}%
are denoted as \textit{structure functions}, where $\alpha _{\ast s}$
identifies the anisotropy-strength function. As shown below, the structure
functions are related to corresponding observable fluid fields. In order for 
$f_{\ast s}$ to be an adiabatic invariant, in the following the structure
functions are required to have a suitable constant value, so that $\Lambda
_{\ast s}=const.$ In the inertial laboratory frame the KDF $f_{\ast \mathrm{s%
}}$ given by (\ref{f*}) becomes%
\begin{eqnarray}
f_{\ast \mathrm{s-lab}} &=&N_{\ast s}\exp \left[ -\gamma _{\ast s}\left(
P_{t}U^{t}+P_{z}U^{z}\right) -\alpha _{\ast s}m^{\prime }\right]  \notag \\
&=&N_{\ast s}\exp \left[ -\gamma _{\ast s}\left( P_{t}\sqrt{1+U^{z2}}%
+P_{z}U^{z}\right) \right]  \notag \\
&&\times \exp \left[ -\alpha _{\ast s}m^{\prime }\right] .
\end{eqnarray}%
For comparison, the expression of the scalar $f_{\ast \mathrm{s}}$ can be
evaluated also in the fluid co-moving frame, in which by definition the
fluid $4-$velocity is $U_{\text{com}}^{\mu }=\left( 1,0,0,0\right) $. Since
the stationarity condition holds also in such a frame, then $f_{\ast \mathrm{%
s}}$ remains an adiabatic invariant and takes the form%
\begin{equation}
f_{\ast \mathrm{s-com}}=N_{\ast s}\exp \left[ -\gamma _{\ast s}P_{t}-\alpha
_{\ast s}m^{\prime }\right] .  \label{eq-com}
\end{equation}

Before concluding it is worth commenting on the validity of the time and $z$%
\ symmetries in the inertial laboratory frame and the fluid co-moving frame.
By definition, the co-moving frame is characterized by 4-velocity $U_{\text{%
com}}^{\mu }=\left( 1,0,0,0\right) $. When seen from the inertial laboratory
frame, the latter takes the form $U^{\mu }=\left( U^{t},0,0,U^{z}\right) $
and is such that $U^{\mu }=U^{\mu }\left( \varepsilon ^{k}r\right) $. It
follows that, in the inertial laboratory frame, the fluid is characterized
by a nearly-constant 4-velocity, where time and spatial variations are
allowed only as a consequence of the \textquotedblleft
slow\textquotedblright\ dynamics implied by the RR effect. Hence, to
leading-order also the co-moving frame is inertial and is moving with
constant spatial velocity in the $z$-direction with respect to the
laboratory frame. In this approximation, the two reference frames are
therefore related by a Lorentz boost with constant velocity. By
construction, a Lorentz boost of this kind does not involve coordinate
transformations, with the result that the two coordinate symmetries with
respect to the time and $z$ components hold the same in the two reference
frames.

\section{Perturbative theory and Chapman-Enskog representation}

In this section we develop a perturbative theory appropriate for the
analytical treatment of the functional dependences and non-isotropic
contributions carried by the equilibrium KDF. This is performed as a
preliminary analysis in order to allow for an analytical calculation of the
corresponding relevant equilibrium fluid fields, to be carried out in
subsequent sections. From the conceptual point of view, the perturbative
theory reported below is based on previous treatments holding for
non-relativistic plasmas \cite{Loop}. Accordingly, in the present
relativistic approach, the perturbative theory can be equivalently carried
out in the inertial laboratory or co-moving frames, with the two being
related by a well-defined Lorentz boost. For convenience, in view of the
following calculation of the equilibrium fluid fields, we shall perform the
perturbative analysis in the co-moving frame. Thus, in terms of the
dimensionless parameter $\varepsilon $, the solution (\ref{f*}) is required
to admit a Chapman-Enskog asymptotic representation of the type%
\begin{equation}
f_{\ast \mathrm{s}}=f_{J\mathrm{s}}\left[ 1+\varepsilon \delta f_{\mathrm{s}%
}+O\left( \varepsilon ^{2}\right) \right] ,  \label{ch-1}
\end{equation}%
where $f_{J\mathrm{s}}$ and $\delta f_{\mathrm{s}}$ are respectively the
leading-order and the first-order (i.e., $O\left( \varepsilon \right) $)
contributions. For clarity, we have left the symbol $\varepsilon $ in front
of $\delta f_{\mathrm{s}}$ to emphasize this is a linear term in the same
expansion parameter.

Let us consider first the leading-order term $f_{J\mathrm{s}}$. This is
identified here with the relativistic Juttner distribution function (also
known as relativistic Maxwellian distribution). Following the notation
adopted in \cite{degroot}, the latter is written in covariant form as%
\begin{equation}
f_{J\mathrm{s}}=\frac{1}{\left( 2\pi \hbar \right) ^{3}}\exp \left[ -\frac{%
P_{\mu }U^{\mu }}{T}\right] ,  \label{Juttner}
\end{equation}%
where\ $P_{\mu }U^{\mu }=P_{t}$ in the fluid co-moving frame and $T$ is
denoted as Juttner temperature. This plays the role analogous of the scalar
temperature in non-relativistic plasmas described by a Maxwellian KDF. The
connection between Eq.(\ref{Juttner}) and Eq.(\ref{eq-com}) is then readily
obtained upon identifying%
\begin{eqnarray}
N_{\ast s} &=&\frac{1}{\left( 2\pi \hbar \right) ^{3}}, \\
\gamma _{\ast s} &=&\frac{1}{T}.
\end{eqnarray}%
As such, the Chapman-Enskog series sought here amounts to effectively
developing an asymptotic representation of the equilibrium KDF that permits
the analytical description of nearly-Maxwellian (in relativistic sense)
plasmas. The deviation from the purely-Maxwellian and isotropic case is
associated here with the linear $O\left( \varepsilon \right) $\ contribution 
$\delta f_{\mathrm{s}}$ that gives rise to non-ideal fluid effects.

Let us now discuss the meaning of the first-order term $\delta f_{\mathrm{s}%
} $. According to the present theoretical framework, the latter contribution
should permit an unambiguous interpretation of the physical meaning of
phase-space anisotropies, to be associated accordingly with the explicit
functional dependence of the equilibrium KDF on the guiding-center particle
magnetic moment through the exponential contribution on $\alpha _{\ast
s}m^{\prime }$. In order to single out in the Chapam-Enskog series the
non-isotropic features associated with $\alpha _{\ast s}m^{\prime }$\ with
respect to the isotropic Juttner solution, the same contribution must be
suitably small compared to the leading-order term. Therefore, in the
co-moving frame we must require that for thermal particles, namely particles
having thermal velocity $v_{th}\sim \sqrt{\frac{T}{m}}$, with $T=const.$
being the Juttner temperature, the following additional ordering can be set:%
\begin{equation}
\left. \alpha _{\ast s}m^{\prime }\right\vert _{v\sim v_{th}}\sim O\left(
\varepsilon \right) .  \label{m'-epsilon}
\end{equation}%
In addition, in order to express the KDF at particle position, an inverse
gyrokinetic transformation must be applied to the guiding-center expression
of the magnetic moment given by Eqs.(\ref{m'})-(\ref{mu'}). On the other
hand, thanks to the ordering (\ref{m'-epsilon}) we can safely write%
\begin{equation}
m^{\prime }=\mu \left[ 1+O\left( \varepsilon \right) \right] ,
\end{equation}%
where now%
\begin{equation}
\mu =\frac{w^{2}}{2qH}
\end{equation}%
represents the leading-order representation of the magnetic moment evaluated
at the particle position. As a consequence, we can finally write%
\begin{equation}
\delta f_{\mathrm{s}}=-\alpha _{\ast s}\frac{w^{2}}{2qH}+O\left( \varepsilon
^{2}\right) ,  \label{delta-f}
\end{equation}%
where $w^{2}$ is the squared of the particle space-like velocity components
in the plane orthogonal to the local magnetic field direction in the given
coordinate system. It is important to stress that in such a picture the
physical meaning of the constant function $\alpha _{\ast s}$ is that it
measures the strength of the anisotropy effect over the isotropic
leading-order solution. The representation (\ref{ch-1})\ is therefore
particularly convenient to isolate the source of explicit phase-space
anisotropy expressed by $\alpha _{\ast s}$.

In conclusion, collecting the previous results, we can write the
Chapman-Enskog series representation of the equilibrium KDF $f_{\ast \mathrm{%
s}}$ given by Eq.(\ref{f*}) in the final form%
\begin{equation}
f_{\ast \mathrm{s}}=N_{\ast s}\exp \left[ -\frac{P_{\mu }U^{\mu }}{T}\right] %
\left[ 1-\frac{\alpha _{\ast s}}{2qH}w^{2}\right] +O\left( \varepsilon
^{2}\right) .  \label{f*-CE}
\end{equation}%
We notice the following issues:

1)\ The advantage of having a leading-order Juttner or relativistic
Maxwellian KDF lies in the possibility of concrete application of the theory
to predictions of physical states associated with either collisionless
plasmas or nearly-Maxwellian collisional plasmas. The latter must be
intended as systems that are allowed to exhibit weak deviations from the
purely Maxwellian collisional case, to occur on time-scales shorter that the
characteristic collisional time-scale of the system. In this sense, although
the theory is performed here for collisionless plasmas, the form of the
Chapman-Enskog series (\ref{ch-1}) with (\ref{Juttner}) can be also applied
to describe collisional plasmas whose KDF exhibits weak deviations from the
equilibrium Juttner distribution.

2) The perturbative theory proposed here permits the analytical treatment of
weakly non-isotropic relativistic plasmas, whereby the source of temperature
and pressure anisotropies is singled out in the $O\left( \varepsilon \right) 
$ polynomial term of the Chapman-Enskog series. The theory then relates the
features of the temperature and pressure anisotropies to the quantity $%
\alpha _{\ast s}$, which is a physical observable (i.e., a continuum fluid
field)\ that enters the definition of the non-ideal contributions to the
pressure tensor. Accordingly, we refer to $\alpha _{\ast s}=const.$ as the
anisotropy-strength coupling constant.

3)\ The identification of the leading-order KDF with the Juttner
distribution is a requirement for the subsequent proof of validity of the
polytropic representation of the EoS for the non-ideal plasma described by
the equilibrium KDF $f_{\ast s}$ given by Eq.(\ref{f*-CE}).

\subsection{Magnetic field structure}

The Chapman-Enskog representation given above is instrumental for the
subsequent analytical calculation of the relevant fluid fields associated
with the kinetic equilibrium. The leading-order solution $f_{J\mathrm{s}}$
is isotropic on particle velocity, namely, according to the definitions
given in Section II, it depends only on $u^{t}$. On the other hand, the $%
O\left( \varepsilon \right) $ correction $\delta f_{\mathrm{s}}$ is
non-isotropic since it carries a dependence on $w^{2}$, namely the square of
the particle velocity components in the plane orthogonal to the local
direction of magnetic field, to be specified in a given reference frame. In
order to express the physical content of the equilibrium KDF we must
characterize the non-ideal fluid solution in terms of precise fluid fields
associated with physical observables of the system. For the following
analytical calculations, this amounts to specifying the structure of the
magnetic field, which in turn allows one to express consistently the
particle velocity components. In the framework of an exact kinetic solution,
one should consider the whole solution for the EM $4-$potential given by
Maxwell equations and expressed as in Eq.(\ref{A-total}).

However, consistent with the perturbative theory developed here, in order to
exemplify the application of the solution, one can envisage suitable
asymptotic orderings to be imposed on the relative contributions of the EM $%
4-$potential, thus permitting the analytical treatment of the solution.
Following the discussion reported in Section IV, two relevant orderings can
be considered, which correspond respectively to the case of dominant
magnetic field self-generated by plasma currents or the case of dominant
externally-generated magnetic field. For the scope of the present research,
we consider in the following the latter case provided by the orderings (\ref%
{e1})-(\ref{e3}). In detail, for a dominant external magnetic field, when
evaluated in the fluid co-moving frame, we therefore assume that $\mathbf{B}%
^{self}\ll \mathbf{B}^{ext}$. Furthermore, introducing local spatial
cylindrical coordinates $(R,\phi ,z)$, in the same domain the magnetic field
is taken to be purely vertical and uniform, namely such that%
\begin{equation}
\mathbf{B}^{ext}=(0,0,B_{z}\equiv const.).  \label{bext-1}
\end{equation}%
This assumption appears appropriate in order to grasp the fundamental
physical properties of the non-ideal fluid solution. As a result of such
topology for the magnetic field, the particle velocity can be conveniently
represented in polar coordinates as $\mathbf{u}=\left( w\cos \phi ,w\sin
\phi ,u_{z}\right) $, which permits a clear identification of the square
term $w^{2}$ in the magnetic moment expression.

\section{Fluid $4-$flow and density}

In this section we compute the fluid $4-$flow $N^{\mu }=\left( n,\mathbf{j}%
\right) $ associated with the kinetic equilibrium solution (\ref{f*-CE}).
Again, for simplicity of notation, we omit the species subscript "s" and we
set for the purpose of the calculation $c=1$. By definition, the
time-component of $N^{\mu }$ yields the fluid number density $n$, while
spatial components provide the fluid velocity. In the following, the
calculation of the fluid fields is carried out in the fluid co-moving frame.
This implies that $N_{\mathrm{com}}^{\mu }=\left( n,\mathbf{0}\right) $, so
that its evaluation brings information on number density, while the fluid
velocity is assumed to be assigned and to coincide with the flow velocity of
the Lorentz transformation relating laboratory and co-moving frames.

We expect that, since the kinetic equilibrium (\ref{f*-CE}) carries a
velocity-space anisotropy and deviates from the Juttner solution, also the
fluid number density retains information about this peculiar structure. More
precisely, for flat tangent space-time where $\sqrt{-g}=1$ and when the
kinematic constraint on the normalization of the particle $4-$velocity
applies, the equilibrium $4-$flow represented by $f_{\ast \mathrm{s}}$ is
defined as%
\begin{equation}
N^{\mu }=\int_{\mathcal{V}^{4}}d^{4}u\Theta \left( u^{0}\right) \delta
\left( u^{\mu }u_{\mu }-1\right) u^{\mu }f_{\ast \mathrm{s}}.
\end{equation}%
After expressing the constraints carried by the theta-function and the
Dirac-delta, it reduces to%
\begin{equation}
N^{\mu }=\int_{\mathcal{V}^{3}}\frac{d^{3}u}{u^{t}}u^{\mu }f_{\ast \mathrm{s}%
},
\end{equation}%
where the time-component of the velocity is $u^{t}=\sqrt{1+\mathbf{u}^{2}}$
and $\mathbf{u}$ identifies the $3-$vector particle velocity. As a
straightforward result, for symmetry reasons on the extrema of integration,
in the co-moving frame the previous expression must be evaluated only for $%
u^{t}$ and it reduces to%
\begin{equation}
N_{\mathrm{com}}^{\mu }=\int_{\mathcal{V}^{3}}d^{3}uf_{\ast \mathrm{s}},
\end{equation}%
in which the dependence carried by $f_{\ast \mathrm{s}}$ on $u^{t}$ are
expressed by means of the kinematic constraint $u^{t}=\sqrt{1+\mathbf{u}^{2}}
$. Substituting the Chapman-Enskog series representation for $f_{\ast 
\mathrm{s}}$ obtained in previous section yields, correct through $O\left(
\varepsilon ^{2}\right) $,%
\begin{equation}
N_{\mathrm{com}}^{\mu }=\int_{\mathcal{V}^{3}}d^{3}uf_{J\mathrm{s}}\left[
1+\delta f_{\mathrm{s}}\right] ,
\end{equation}%
with $f_{J\mathrm{s}}$ and $\delta f_{\mathrm{s}}$ being expressed
respectively by (\ref{Juttner}) and (\ref{delta-f}). This permits to write
the number density in series expansion correspondingly as%
\begin{equation}
n=n_{J}+\delta n,
\end{equation}%
where it is understood that $n_{J}$ is given solely by the equilibrium
Juttner distribution, and it represents the relativistic Maxwellian
contribution, while $\delta n$ arises due to kinetic anisotropy term. We now
evaluate the two contributions separately, providing the mathematical
details of the calculations in order to illustrate the analytical procedure.
This analysis is instrumental for the subsequent calculations involving the
pressure tensor.

We start from the leading-order Juttner density $n_{J}$:%
\begin{equation}
n_{J}=\int_{\mathcal{V}^{3}}d^{3}uf_{J\mathrm{s}}=N_{\ast s}\int_{\mathcal{V}%
^{3}}d^{3}ue^{-\frac{P_{\mu }U^{\mu }}{T}},
\end{equation}%
where in the co-moving frame $P_{\mu }U_{\text{com}}^{\mu }=P_{t}$ and $%
P^{t}=(u^{t}+qA^{t})=(u^{t}+q\Phi )$, with $\Phi $ denoting the
electrostatic potential. Replacing this representation and adopting
spherical coordinates in velocity space we obtain%
\begin{equation}
n_{J}=N_{\ast s}4\pi e^{-\frac{q\Phi }{T}}\int_{0}^{+\infty }due^{-\frac{%
\sqrt{1+u^{2}}}{T}}u^{2}.
\end{equation}%
The integral can be evaluated analytically by means of change of variables
introducing the quantities $y\equiv \frac{1}{T}$ and $\tau \equiv \frac{%
\sqrt{1+u^{2}}}{T}$, from which one has the corresponding differential
identity $du=T\tau \left( \tau ^{2}-y^{2}\right) ^{-1/2}d\tau $. This gives%
\begin{equation}
n_{J}=N_{\ast s}4\pi T^{3}e^{-\frac{q\Phi }{T}}\int_{y}^{+\infty }d\tau
\left( \tau ^{2}-y^{2}\right) ^{1/2}\tau e^{-\tau },
\end{equation}%
which can be analytically solved and expressed in terms of modified Bessel
function of second kind \cite{degroot}%
\begin{equation}
K_{n}\left( y\right) =\frac{2^{n}n!}{\left( 2n\right) !}\frac{1}{y^{n}}%
\int_{y}^{+\infty }d\tau \left( \tau ^{2}-y^{2}\right) ^{n-1/2}e^{-y},
\label{B1}
\end{equation}%
which can be transformed by partial integration in the equivalent form%
\begin{equation}
K_{n}\left( y\right) =\frac{2^{n-1}\left( n-1\right) !}{\left( 2n-2\right) }%
\frac{1}{y^{n}}\int_{y}^{+\infty }d\tau \left( \tau ^{2}-y^{2}\right)
^{n-3/2}\tau e^{-\tau }.  \label{B2}
\end{equation}%
The final result for the Juttner density depends on the Bessel function of
order 2, namely $K_{2}\left( \frac{1}{T}\right) $, and is found to be%
\begin{equation}
n_{J}=N_{\ast s}4\pi Te^{-\frac{q\Phi }{T}}K_{2}\left( \frac{1}{T}\right) .
\label{n_J}
\end{equation}

Let us now consider the linear contribution $\delta n$, which carries the
non-isotropic velocity dependence on the component $w^{2}$. This is defined
by the integral%
\begin{equation}
\delta n=\int_{\mathcal{V}^{3}}d^{3}uf_{J\mathrm{s}}\delta f_{\mathrm{s}%
}=N_{\ast s}\int_{\mathcal{V}^{3}}d^{3}u\left( -\frac{\alpha _{\ast s}}{2qH}%
w^{2}\right) e^{-\frac{P_{\mu }U^{\mu }}{T}},
\end{equation}%
where the magnetic field is taken according to the ordering introduced
above, so that here $H=B_{z}$ in the fluid co-moving frame. Given the
non-isotropic functional dependence of the expression on particle velocity,
contrary to the isotropic Juttner term, in this case the integral can be
conveniently calculated introducing polar coordinates $\left( w\cos \phi
,w\sin \phi ,u_{z}\right) $, namely%
\begin{equation}
\int d^{3}u\rightarrow \int_{0}^{2\pi }d\phi \int_{0}^{+\infty
}wdw\int_{-\infty }^{+\infty }du_{z},
\end{equation}%
so that $u^{2}=w^{2}+u_{z}^{2}$. Performing the integral over $d\phi $, the
density correction $\delta n$ then becomes%
\begin{equation}
\delta n=-\frac{\alpha _{\ast s}}{2qH}N_{\ast s}2\pi e^{-\frac{q\Phi }{T}%
}\int_{0}^{+\infty }w^{3}dw\int_{-\infty }^{+\infty }e^{-\frac{\sqrt{%
1+w^{2}+u_{z}^{2}}}{T}}du_{z}.
\end{equation}%
To compute the integral, we introduce the following parametrization:%
\begin{eqnarray}
\gamma  &\equiv &\sqrt{1+w^{2}+u_{z}^{2}}, \\
\gamma _{z} &\equiv &\sqrt{1+u_{z}^{2}},
\end{eqnarray}%
and then the change of variables by setting%
\begin{eqnarray}
x &\equiv &\frac{1}{T}\left( \gamma -\gamma _{z}\right) , \\
y &\equiv &\frac{1}{T}\gamma _{z}.
\end{eqnarray}%
After some algebraic simplifications one obtains the expression%
\begin{eqnarray}
\delta n &=&-\frac{\alpha _{\ast s}}{2qH}N_{\ast s}4\pi T^{5}e^{-\frac{q\Phi 
}{T}}  \notag \\
&&\times \int_{0}^{+\infty }dx\int_{1/T}^{+\infty }dy\left[ \mathbf{I+II+III}%
\right] ,
\end{eqnarray}%
which is found to be composed by the sum of three separate integrals, where
respectively%
\begin{eqnarray}
\mathbf{I} &=&\left( x^{3}e^{-x}\right) y\left( y^{2}-\frac{1}{T^{2}}\right)
^{-1/2}e^{-y}, \\
\mathbf{II} &=&3\left( x^{2}e^{-x}\right) y^{2}\left( y^{2}-\frac{1}{T^{2}}%
\right) ^{-1/2}e^{-y}, \\
\mathbf{III} &=&2\left( xe^{-x}\right) y^{3}\left( y^{2}-\frac{1}{T^{2}}%
\right) ^{-1/2}e^{-y}.
\end{eqnarray}%
These three contributions can be solved analytically and expressed again in
terms of the modified Bessel functions of second type $K_{n}\left( y\right) $
according to Eqs.(\ref{B1}) and (\ref{B2}). In detail, we notice that in the
previous expressions the functional dependence in terms of integration
variables $x$\ and $y$\ is decoupled. Thus, the integrals over variable $x$\
can be solved thanks to the following analytical representation%
\begin{equation}
\int_{0}^{+\infty }x^{n}e^{-ax}dx=\frac{n!}{a^{n+1}},
\end{equation}%
where the variable $a$\ in the exponential is such that $a=1$\ identically
in all cases considered here. Instead, concerning the integration over
variable $y$, each expression must be ultimately reduced to either Eqs.(\ref%
{B1}) and (\ref{B2}). In this respect we notice that the following identity
holds%
\begin{equation}
y^{2}\left( y^{2}-\frac{1}{T^{2}}\right) ^{-1/2}e^{-y}=ye^{-y}\frac{d}{dy}%
\left[ \left( y^{2}-\frac{1}{T^{2}}\right) ^{1/2}\right] ,
\end{equation}%
which can be iterated in combination with integration by parts in order to
lower the power of $y$ until the desired representation (\ref{B1}) or (\ref%
{B2}) is reached. This yields explicitly the solutions%
\begin{eqnarray}
\mathbf{I} &=&\frac{6}{T}K_{1}\left( \frac{1}{T}\right) , \\
\mathbf{II} &=&-\frac{6}{T}K_{1}\left( \frac{1}{T}\right) +6\frac{1}{T^{2}}%
K_{2}\left( \frac{1}{T}\right) , \\
\mathbf{III} &=&-\frac{14}{3}\frac{1}{T^{2}}K_{2}\left( \frac{1}{T}\right) +%
\frac{2}{9}\frac{1}{T^{3}}K_{3}\left( \frac{1}{T}\right) .
\end{eqnarray}%
Collecting the three terms and making use of the relation%
\begin{equation}
K_{n+1}\left( y\right) =2n\frac{K_{n}\left( y\right) }{y}+K_{n-1}\left(
y\right) ,  \label{prop}
\end{equation}%
the perturbative density follows:%
\begin{equation}
\delta n=-\frac{\alpha _{\ast s}}{2qH}\left[ \frac{20}{9}T^{2}+\frac{2}{9}T%
\frac{K_{1}\left( \frac{1}{T}\right) }{K_{2}\left( \frac{1}{T}\right) }%
\right] n_{J}.
\end{equation}%
Hence, the equilibrium number density corresponding to the non-ideal
relativistic plasma fluid described by the equilibrium KDF (\ref{f*}),
correct through $O\left( \varepsilon ^{2}\right) $ becomes%
\begin{equation}
n=n_{J}\left[ 1-\frac{\alpha _{\ast s}}{2qH}\left( \frac{20}{9}T^{2}+\frac{2%
}{9}T\frac{K_{1}\left( \frac{1}{T}\right) }{K_{2}\left( \frac{1}{T}\right) }%
\right) \right] .
\end{equation}

The present calculation shows how the velocity-space anisotropy of the
equilibrium KDF ultimately affects the representation for the non-ideal
fluid number density with respect to the Juttner isotropic solution. The
result can be obtained in closed analytical form thanks to the
Chapman-Enskog representation in which the anisotropic term enters as a
polynomial weighted over the Juttner exponential factor.

\section{Fluid stress-energy tensor}

In this section we proceed with the analytical calculation of the
stress-energy $4-$tensor associated with the kinetic equilibrium (\ref{f*}).
This tensor brings information about the thermodynamical properties of the
corresponding non-ideal plasma fluid, whereby non-ideal features arise due
to the occurrence of the phase-space anisotropy in the KDF. The expression
of the stress-energy tensor is demanded in order to discuss the resulting
equation-of-state properties of non-ideal plasma jets under the action of
pressure anisotropy. Thus, invoking the definition given above by Eq.(\ref%
{tmunu-bis}), and considering again a flat space-time in the tangent
velocity space, we write the relativistic equilibrium stress-energy tensor as%
\begin{equation}
T^{\mu \nu }=\int_{\mathcal{V}^{3}}\frac{d^{3}u}{\sqrt{1+u^{2}}}u^{\mu
}u^{\nu }f_{\ast \mathrm{s}}.
\end{equation}%
In view of the Chapman-Enskog series representation for $f_{\ast \mathrm{s}}$
obtained above in Eq.(\ref{ch-1}), correct through $O\left( \varepsilon
_{s}^{2}\right) $ the same tensor takes the form%
\begin{equation}
T^{\mu \nu }=\int_{\mathcal{V}^{3}}\frac{d^{3}u}{\sqrt{1+u^{2}}}u^{\mu
}u^{\nu }f_{J\mathrm{s}}\left[ 1+\delta f_{\mathrm{s}}\right] .
\end{equation}%
As for the fluid $4-$flow, because of the mathematical structure of the
kinetic solution, also in this case the tensor is composed of two
contributions, namely the leading-order term associated with the Juttner
(i.e., relativistic Maxwellian) function $f_{J\mathrm{s}}$ and the
perturbative correction due to $\delta f_{\mathrm{s}}$, which expresses the
non-ideal character of the fluid solution. Hence, we can conveniently
decompose the tensor $T^{\mu \nu }$ as%
\begin{equation}
T^{\mu \nu }=T_{J}^{\mu \nu }+\delta \pi ^{\mu \nu },  \label{T-decomposed}
\end{equation}%
where the subscript $"J"$ stands for Juttner term and the meaning of the
remaining symbols is understood.

We consider the calculation of the two contributions to $T^{\mu \nu }$
separately, starting from the leading-order Juttner tensor. In this case the
solution takes the well-known representation given above by Eq.(\ref{T-max}%
), namely%
\begin{equation}
T_{J}^{\mu \nu }=n_{J}e_{J}U^{\mu }U^{\nu }-P\Delta ^{\mu \nu },  \label{kkk}
\end{equation}%
where $n_{J}$ is the equilibrium Juttner particle number density (\ref{n_J})
and $e_{J}$ is the Juttner energy per particle, while $P$ is the scalar
isotropic pressure and $\Delta ^{\mu \nu }=g^{\mu \nu }-U^{\mu }U^{\nu }$.
One has in particular that (setting $m_{o}=c=1$)%
\begin{equation}
e_{J}=\frac{K_{1}\left( \frac{1}{T}\right) }{K_{2}\left( \frac{1}{T}\right) }%
+3T,  \label{ej}
\end{equation}%
while the scalar pressure is%
\begin{equation}
P=n_{J}T,
\end{equation}%
which recovers the equation of state of an ideal Maxwellian plasma fluid,
consistently with the Chapman-Enskog approximation scheme.

Let us now evaluate the linear contribution $\delta \pi ^{\mu \nu }$:%
\begin{equation}
\delta \pi ^{\mu \nu }=\int_{\mathcal{V}^{3}}\frac{d^{3}u}{\sqrt{1+u^{2}}}%
u^{\mu }u^{\nu }\delta f_{\mathrm{s}}f_{J\mathrm{s}},
\end{equation}%
namely%
\begin{equation}
\delta \pi ^{\mu \nu }=N_{\ast s}\int_{\mathcal{V}^{3}}\frac{d^{3}u}{\sqrt{%
1+u^{2}}}u^{\mu }u^{\nu }\left( -\frac{\alpha _{\ast s}}{2qH}w^{2}\right)
e^{-\frac{P_{\mu }U^{\mu }}{T}}.
\end{equation}%
It is clear from symmetry arguments that the only nonvanishing contributions
to the tensor $\delta \pi ^{\mu \nu }$ are the diagonal terms, so that $%
\delta \pi ^{\mu \nu }$ is a diagonal tensor. Again, because of the chosen
structure of purely-vertical magnetic field in the fluid co-moving frame,
the velocity integral is conveniently represented in polar coordinates with $%
\mathbf{u}=\left( w\cos \phi ,w\sin \phi ,u_{z}\right) $. Hence, $\delta \pi
^{\mu \nu }$ reduces to%
\begin{eqnarray}
\delta \pi ^{\mu \nu } &=&-\frac{\alpha _{\ast s}}{2qH}N_{\ast s}e^{-\frac{%
q\Phi }{T}}\int_{0}^{2\pi }d\phi \int_{0}^{+\infty }w^{3}dw  \notag \\
&&\times \int_{-\infty }^{+\infty }du_{z}\frac{u^{\mu }u^{\nu }}{\sqrt{%
1+u^{2}}}e^{-\frac{\sqrt{1+w^{2}+u_{z}^{2}}}{T}}.  \label{delta-pai}
\end{eqnarray}%
We start from the time-time component $\delta \pi ^{tt}$, obtained for the
dyadic $u^{t}u^{t}$. Invoking the kinematic constraint holding for the same
time-component $u^{t}$ and performing the integral over $d\phi $, after
algebraic simplification one obtains 
\begin{eqnarray}
\delta \pi ^{tt} &=&-\frac{\alpha _{\ast s}}{2qH}N_{\ast s}2\pi e^{-\frac{%
q\Phi }{T}}\int_{0}^{+\infty }w^{3}dw  \notag \\
&&\times \int_{-\infty }^{+\infty }du_{z}\sqrt{1+u^{2}}e^{-\frac{\sqrt{%
1+w^{2}+u_{z}^{2}}}{T}}.
\end{eqnarray}%
The integral can be evaluated analytically performing the same type of
mathematical steps outlined in the previous section for the fluid number
density. By iteration of the procedure and by invoking the definitions of
modified Bessel functions finally yields the expression 
\begin{eqnarray}
\delta \pi ^{tt} &=&-\frac{\alpha _{\ast s}}{2qH}N_{\ast s}4\pi e^{-\frac{%
q\Phi }{T}}T^{4}  \notag \\
&&\left[ \frac{8}{3}K_{2}\left( \frac{1}{T}\right) -\frac{2}{T}K_{3}\left( 
\frac{1}{T}\right) +\frac{2}{T^{2}}K_{4}\left( \frac{1}{T}\right) \right] 
\end{eqnarray}%
Invoking the solution for the Juttner density gives the compact form%
\begin{equation}
\delta \pi ^{tt}=-\frac{\alpha _{\ast s}}{2qH}\left[ \frac{8}{3}-\frac{2}{T}%
\frac{K_{3}\left( \frac{1}{T}\right) }{K_{2}\left( \frac{1}{T}\right) }+%
\frac{2}{T^{2}}\frac{K_{4}\left( \frac{1}{T}\right) }{K_{2}\left( \frac{1}{T}%
\right) }\right] T^{3}n_{J}.  \label{d-tt}
\end{equation}%
Let us now consider the component $\delta \pi ^{zz}$. Again, from Eq.(\ref%
{delta-pai}) after performing the integral over $d\phi $ we obtain%
\begin{eqnarray}
\delta \pi ^{zz} &=&-\frac{\alpha _{\ast s}}{2qH}N_{\ast s}2\pi e^{-\frac{%
q\Phi }{T}}\int_{0}^{+\infty }w^{3}dw  \notag \\
&&\times \int_{-\infty }^{+\infty }du_{z}\frac{u_{z}^{2}}{\sqrt{1+u^{2}}}e^{-%
\frac{\sqrt{1+w^{2}+u_{z}^{2}}}{T}}.
\end{eqnarray}%
The analytical calculation yields%
\begin{eqnarray}
\delta \pi ^{zz} &=&-\frac{\alpha _{\ast s}}{2qH}N_{\ast s}4\pi e^{-\frac{%
q\Phi }{T}}T^{4}  \notag \\
&&\left[ \frac{4}{3}K_{2}\left( \frac{1}{T}\right) +\frac{2}{9}\frac{1}{T}%
K_{3}\left( \frac{1}{T}\right) \right] ,
\end{eqnarray}%
which can be written in terms of the Juttner pressure $P=n_{J}T$ as%
\begin{equation}
\delta \pi ^{zz}=-\frac{\alpha _{\ast s}}{2qH}T^{2}\left[ \frac{4}{3}+\frac{2%
}{9}\frac{1}{T}\frac{K_{3}\left( \frac{1}{T}\right) }{K_{2}\left( \frac{1}{T}%
\right) }\right] P.  \label{dp-zz}
\end{equation}%
Finally, let us evaluate the two components $\delta \pi ^{xx}$ and $\delta
\pi ^{yy}$. For the geometry configuration of the problem considered here,
the two contributions to the stress-energy tensor are the same, namely $%
\delta \pi ^{xx}=\delta \pi ^{yy}$. This feature can be easily verified by
using polar coordinates, so that $u_{x}^{2}=w^{2}\cos ^{2}\phi $ and $%
u_{y}^{2}=w^{2}\sin ^{2}\phi $, and noting that $\int_{0}^{2\pi }\cos
^{2}\phi d\phi =\int_{0}^{2\pi }\sin ^{2}\phi d\phi =\pi $. We calculate $%
\delta \pi ^{xx}$. From Eq.(\ref{delta-pai}) we then obtain%
\begin{eqnarray}
\delta \pi ^{xx} &=&-\frac{\alpha _{\ast s}}{2qH}\pi N_{\ast s}e^{-\frac{%
q\Phi }{T}}\int_{0}^{+\infty }w^{5}dw  \notag \\
&&\times \int_{-\infty }^{+\infty }du_{z}\frac{1}{\sqrt{1+u^{2}}}e^{-\frac{%
\sqrt{1+w^{2}+u_{z}^{2}}}{T}}.
\end{eqnarray}%
Explicit calculation provides the following analytical solution:%
\begin{eqnarray}
\delta \pi ^{xx} &=&-\frac{\alpha _{\ast s}}{2qH}2\pi N_{\ast s}e^{-\frac{%
q\Phi }{T}}T^{4}  \notag \\
&&\left[ \frac{16}{3}K_{2}\left( \frac{1}{T}\right) +\frac{8}{9}\frac{1}{T}%
K_{3}\left( \frac{1}{T}\right) \right] ,
\end{eqnarray}%
which again is conveniently expressed in terms of the Juttner pressure $%
P=n_{J}T$ as%
\begin{equation}
\delta \pi ^{xx}=-\frac{\alpha _{\ast s}}{2qH}T^{2}\left[ \frac{8}{3}+\frac{4%
}{9}\frac{1}{T}\frac{K_{3}\left( \frac{1}{T}\right) }{K_{2}\left( \frac{1}{T}%
\right) }\right] P.  \label{dp-xx}
\end{equation}%
We notice that from the previous results the identity $\delta \pi
^{xx}=2\delta \pi ^{zz}$ follows.

This completes the calculation of the components of the stress-energy tensor
for the relativistic non-ideal plasma fluid, evaluated in the fluid
co-moving frame. In the following, two physical applications are considered,
which follow from the knowledge of the stress-energy tensor and its
non-isotropic character. These concern, respectively, the evaluation of the
energy density of the non-ideal plasma, which is provided by the time-time
component of $T^{\mu \nu }$\ in the co-moving frame, and the representation
of the pressure tensor, which is instead identified with the space-space
components of $T^{\mu \nu }$\ in the same frame.

\subsection{Application: the fluid energy density}

Let us consider first the calculation of the fluid energy density $ne$,
which is defined out of the stress-energy tensor as%
\begin{equation}
ne\equiv T^{\mu \nu }U_{\mu }U_{\nu }.
\end{equation}%
Inserting the decomposition of $T^{\mu \nu }$\ according to Eq.(\ref%
{T-decomposed}) and invoking the Juttner solution (\ref{kkk}) gives in the
co-moving frame%
\begin{equation}
ne=n_{J}e_{J}+\delta \pi ^{tt},
\end{equation}%
where $n_{J}$\ is the Juttner density given by Eq.(\ref{n_J}), $e_{J}$\ is
given by Eq.(\ref{ej}) and the analytic expression of $\delta \pi ^{tt}$\ is
provided by Eq.(\ref{d-tt}). Invoking Eq.(\ref{prop}) we can write Eq.(\ref%
{d-tt}) in the equivalent way as%
\begin{equation}
\delta \pi ^{tt}=-\frac{\alpha _{\ast s}}{2qH}\left[ \frac{8}{3}+\frac{10}{T}%
\frac{K_{3}\left( \frac{1}{T}\right) }{K_{2}\left( \frac{1}{T}\right) }+%
\frac{2}{T^{2}}\right] T^{3}n_{J},
\end{equation}%
from which we can infer that necessarily%
\begin{equation}
\delta \pi ^{tt}<0.
\end{equation}%
As a consequence, this permits to prove that the total energy density is
reduced with respect to the purely isotropic configuration corresponding to
the Juttner solution, namely%
\begin{equation}
ne=n_{J}e_{J}-|\delta \pi ^{tt}|.
\end{equation}%
The result is found to be a unique consequence of the macroscopic continuum
fluid effect of the statistical phase-space anisotropy carried by the
equilibrium KDF describing the non-ideal plasma fluid.

\subsection{Application: the fluid pressure tensor}

Given the results obtained above, the purpose of this section is to prove
that the contribution of phase-space anisotropy ultimately generates a
pressure anisotropy, which expresses the non-ideal character of the plasma
fluid. This shows up in the diagonal space-space components of\ $T^{\mu \nu }
$ and is such that $\delta \pi ^{xx}=\delta \pi ^{yy}\neq \delta \pi ^{zz}$.
This means that, in the co-moving frame, the thermodynamical properties of
the fluid along the local direction of magnetic field, i.e. the $z-$%
direction, are decoupled from those in the plane orthogonal to the magnetic
field lines. The deviation from the isotropic Juttner configuration can be
evaluated analytically and the final expression provides at the same time
the reference solution for the physical characterization of the function $%
\alpha _{\ast s}$ in terms of an observable fluid field. More precisely, in
order to summarize the results obtained and in view of the next
developments, it is instructive to represent the pressure-tensor of the
non-ideal fluid, which is identified in the co-moving frame and corresponds
to the space-space section of the stress-energy tensor $T^{\mu \nu }$,
namely for $\mu ,\nu =1,2,3$ (corresponding to spatial components $x,y,z$).
We denote the $3\times 3$ species pressure tensor with the symbol $%
\underline{\underline{\mathbf{\Pi }}}$ identified as follows:%
\begin{equation}
\underline{\underline{\mathbf{\Pi }}}=P\left( \underline{\underline{\mathbf{I%
}}}+\underline{\underline{\mathbf{\delta \pi }}}\right) ,
\end{equation}%
where $\underline{\underline{\mathbf{I}}}$ is the identity matrix and $%
\underline{\underline{\mathbf{\delta \pi }}}$ carries the first-order
diagonal and non-isotropic correction terms. Since $\underline{\underline{%
\mathbf{\delta \pi }}}$ contains only negative contributions, we can write
explicitly the tensor $\underline{\underline{\mathbf{\Pi }}}$ as%
\begin{equation}
\underline{\underline{\mathbf{\Pi }}}=P\left( 
\begin{array}{ccc}
1-2\delta \pi _{a} & 0 & 0 \\ 
0 & 1-2\delta \pi _{a} & 0 \\ 
0 & 0 & 1-\delta \pi _{a}%
\end{array}%
\right) ,  \label{p-tensor}
\end{equation}%
where the expression for $\delta \pi _{a}$\ follows from equations (\ref%
{dp-zz}) and is given by%
\begin{equation}
\delta \pi _{a}=\frac{\alpha _{\ast s}}{2qH}T^{2}\left[ \frac{4}{3}+\frac{2}{%
9}\frac{1}{T}\frac{K_{3}\left( \frac{1}{T}\right) }{K_{2}\left( \frac{1}{T}%
\right) }\right] .
\end{equation}%
We notice that the pressure tensor is affected by the non-ideal
contributions in two different ways. The first one is through a modification
of the number density, while the second one arises through the non-isotropic
temperature, namely a different statistics characterizing the velocity
dispersion in the directions parallel and perpendicular to the local
magnetic field lines. The non-isotropic corrections to the Juttner isotropic
pressure exhibit characteristic dependencies on the temperature and also
configuration-space dependences contained in the magnetic field $H$.
Finally, it must be stressed that the non-isotropic pressure matrix $%
\underline{\underline{\mathbf{\delta \pi }}}$ yields negative contributions
to the leading-order isotropic Juttner pressure $P$. The occurrence of
microscopic phase-space conservation of particle magnetic moment therefore
ultimately acts as a whole in such a way to reduce the Juttner pressure, and
the decrease is notably non-isotropic, namely it exhibits different
magnitudes along spatial directions.

\section{Polytropic form of the pressure tensor}

The analytical solution (\ref{p-tensor}) determined above for the
non-isotropic pressure tensor $\underline{\underline{\mathbf{\Pi }}}$
represents effectively the equation of state (EoS) for the corresponding
equilibrium non-ideal plasma fluid. Based on this result, we now propose a
technique that allows one to reach a polytropic representation for the same
EoS, and therefore for the same tensor $\underline{\underline{\mathbf{\Pi }}}
$. The approach follows an analogous one first proposed for non-relativistic
plasmas in Ref.\cite{F2023}. The final goal is to represent the dependences
on Juttner temperature $T$ appearing in each non-vanishing entry of the
tensor $\underline{\underline{\mathbf{\Pi }}}$ in terms of assigned
functions of mass density $\rho $ with a characteristic power-law
dependence, namely with a precise polytropic index $\Gamma $.

The starting point is the assumption of validity of polytropic EoS for the
Juttner isotropic pressure, namely the relation:%
\begin{equation}
P=n_{J}T\equiv \kappa \rho ^{\Gamma },
\end{equation}%
where $\rho =m_{o}n_{J}\equiv \rho _{J}$ is the Juttner mass density, $%
\kappa $ is a suitable dimensional function of proportionality and $\Gamma $
is the polytropic index. For completeness and to enhance comparison with
previous works, we leave here indicated the dependence on particle mass $%
m_{o}$. Notice that both $\kappa $ and $\Gamma $ are in principle
species-dependent quantities, but nevertheless we omit the subscript "$s$"
for convenience of notation. From the previous equation we can then obtain a
representation of the isotropic temperature as follows%
\begin{equation}
T\equiv \kappa m_{o}\rho ^{\Gamma -1}.
\end{equation}%
We then assume that the same polytropic relationship for the temperature can
be iterated also in the perturbative non-isotropic corrections to the
pressure tensor, namely for $\underline{\underline{\mathbf{\delta \pi }}}$.
This method is allowed and justified in the framework of the perturbative
kinetic theory pursued here. In this way, the corresponding polytropic
representation of the pressure tensor is reached, in which each single
directional pressure depends only on assigned mass-density profile rather
than the combination of density and temperature dependences.

We consider separately the two entries $\delta \pi ^{xx}=\delta \pi ^{yy}$
and $\delta \pi ^{zz}$. The result is as follows:%
\begin{eqnarray}
\delta \pi _{\rho }^{xx} &=&-\frac{\alpha _{\ast s}}{2qH}\kappa \left(
\kappa m_{o}\right) ^{2}\rho ^{3\Gamma -2}  \notag \\
&&\left[ \frac{8}{3}+\frac{4}{9}\frac{\rho ^{1-\Gamma }}{\kappa m_{o}}\frac{%
K_{3}\left( \frac{\rho ^{1-\Gamma }}{\kappa m_{o}}\right) }{K_{2}\left( 
\frac{\rho ^{1-\Gamma }}{\kappa m_{o}}\right) }\right] , \\
\delta \pi _{\rho }^{zz} &=&-\frac{\alpha _{\ast s}}{2qH}\kappa \left(
\kappa m_{o}\right) ^{2}\rho ^{3\Gamma -2}  \notag \\
&&\left[ \frac{4}{3}+\frac{2}{9}\frac{\rho ^{1-\Gamma }}{\kappa m_{o}}\frac{%
K_{3}\left( \frac{\rho ^{1-\Gamma }}{\kappa m_{o}}\right) }{K_{2}\left( 
\frac{\rho ^{1-\Gamma }}{\kappa m_{o}}\right) }\right] ,
\end{eqnarray}%
where we have made use of the subscript "$\rho $" to indicate the polytropic
dependence. We notice that the functional form of both contributions scales
with a precise power-law of the density profile. Analogous expressions can
be readily obtained also for the corresponding quantities $\delta \pi _{a}$
and $\delta \pi _{b}$ in Eq.(\ref{p-tensor}).

The outcome is relevant because it proves that, under validity of
appropriate ordering assumptions, deviations from purely Maxwellian
distribution responsible for the onset of non-isotropic pressure tensor can
still be handled in the framework of a polytropic solution. This provides
corresponding directional equations of state that depend only on plasma
density with specific power-law coefficients. The expression for the
pressure tensor provides at the same time the kinetic closure condition for
the equilibrium hydrodynamic equations describing the non-ideal plasma
fluid. The result can be useful either for interpretation of observational
data or for numerical studies of jets dynamical and thermodynamical
properties. Applications include in particular the case of astrophysical
magnetized plasma jets and gamma-ray burst as well as, in principle,
laboratory laser plasmas.

\section{Conclusions}

The description of relativistic plasma jets forming equilibrium structures
represents an intriguing subject of research for both theoretical and
astrophysical studies. In fact, systems of this type are expected to exhibit
peculiarities that arise as a consequence of microscopic collective features
that require statistical approaches to be exploited and predicted. This
concerns in particular the occurrence of so-called phase-space or kinetic
anisotropies giving rise to the manifestation of pressure anisotropy effects
characterizing corresponding non-ideal continuum fluid states. The issue
ultimately translates into the problem of determining the fluid equation of
state (EoS) which encodes the information about dynamical and
thermodynamical properties of the plasma and provides at the same time the
constituent closure condition for the fluid magnetohydrodynamic equations.
For non-ideal relativistic\ plasma fluids this amounts to replacing the
expression of a single scalar pressure with a non-isotropic pressure tensor
exhibiting different directional pressures. This information is contained in
the stress-energy $4-$tensor, to be evaluated in the fluid co-moving
reference frame. Remarkably,\ in the framework of kinetic theory implemented
here the latter tensor can be calculated consistently as a derived quantity,
namely by means of a suitable average integral of kinetic distribution
function (KDF)\ over particle velocity space.

In order to reach the target, precise conceptual steps have been
consolidated. First, an equilibrium solution for the KDF has been obtained
in analytical form, which is expressed in terms of a Gaussian-like solution
depending only on particle adiabatic invariants characteristic for
relativistic jet plasmas in magnetic fields. In this reference, it has been
shown that the occurrence of phase-space anisotropy generating pressure
anisotropy is associated with the particle magnetic moment conservation.
Second, a perturbative treatment for the equilibrium KDF has been realized
through a Chapman-Enskog expansion carried out around a leading-order
equilibrium and isotropic Juttner (i.e., relativistic Maxwellian)
distribution. As a third target, this polynomial representation for the
equilibrium KDF has allowed for the explicit analytical calculation of the
corresponding stress-energy tensor, evaluated in the fluid co-moving frame,
and the related proof of its non-isotropic character. Finally, the validity
of a polytropic representation for the pressure tensor, with inclusion of
the perturbative non-isotropic contributions, has been proved to hold. This
realizes a non-isotropic tensorial equation of state for the non-ideal jet
plasma fluid which is expressed only as a function of the same fluid mass
density.

The results proposed in the paper are established on a comprehensive
theoretical background and they are expected to have a wide physical
relevance in plasma physics, fluid dynamics and astrophysical plasmas. The
theoretical framework established in the paper can be relevant for the
understanding of equilibrium dynamical and thermodynamical properties of
astrophysical plasmas generating non-ideal fluids belonging to relativistic
jets or gamma-ray bursts as well as for soft-matter studies focused on the
issue of equation of state for complex non-linear statistical systems like
magnetized plasmas. For this reason, the kinetic theory presented in the
paper is susceptible of further investigation and is expected to help
gaining insights into the complex dynamics governing magnetized plasmas and
fluids, with particular reference to relativistic regimes. Thus, the
analytical determination of non-isotropic pressure tensors based on kinetic
approach finds applications in hydrodynamics and magneto-hydrodynamics
studies of non-ideal astrophysical fluids, e.g., through the prescription of
corresponding equations of state and fluid closure conditions. In fact, the
mathematical technique proposed here allows one to overcome the formalism
based on the isotropic Juttner or relativistic Maxwellian kinetic
distribution function describing distribution of velocities of plasma
charges corresponding to ideal plasma fluid state. In particular, it
provides a convenient framework for the investigation of non-ideal physical
phenomena related to the occurrence of microscopic conservation laws,
phase-space anisotropies and the generation of non-isotropic fluid pressures
established through the tensorial representation of the plasma matter
equation of state. Finally, the same theoretical outcome obtained in this
paper is expected to be relevant for future investigations on stability
properties of non-ideal relativistic plasma fluids with plausible
consequences on the improvement of our knowledge of the complex
phenomenology characterizing these systems.

\bigskip

\textbf{CONFLICTS OF INTEREST}

The author has no conflicts to disclose.

\textbf{DATA AVAILABILITY}

The data that support the findings of this study are available within the
article.

\bigskip

\section{Appendix:\ Integrals of motion in relativistic plasmas}

In this Appendix we proceed with the proof about the non-existence of exact
conservation laws in relativistic plasmas, implying non-existence of exact
integrals of motion. We recall that single-particle conservation laws can be
obtained from symmetry properties of the Lagrangian function corresponding
to the particle dynamics. As shown in Refs.\cite{EPJ2,EPJ5}, a consistent
Lagrangian and Hamiltonian descriptions of the EM-RR can be obtained for
finite-size particles modeled in classical picture as having invariant
rest-frame radius $\sigma $ with spherically-symmetric mass and charge
distributions (in particle rest-frame). This is reached by making use of the
non-local effective scalar Lagrangian%
\begin{equation}
L_{eff}\equiv L_{M}(r)+L_{C}^{(ext)}(r)+L_{C}^{(pl)}(r)+2L_{C}^{(RR)}(r, 
\left[ r\right] ),  \label{extremal Lagrangian-0}
\end{equation}%
where $L_{M}(r),$ $L_{C}^{(ext)}(r)$ and $L_{C}^{(pl)}(r)$ are respectively%
\begin{eqnarray}
L_{M}(r) &=&\frac{1}{2}m_{o}c\frac{dr_{\mu }}{ds}\frac{dr^{\mu }}{ds},
\label{LAGRANGIAN -constarint-2} \\
L_{C}^{(ext)}(r) &=&\frac{q}{c}\frac{dr}{ds}^{\mu }\overline{A}_{\mu
}^{(ext)}(r),  \label{LAGRANGIAN -EXTERNAL EM} \\
L_{C}^{(pl)}(r) &=&\frac{q}{c}\frac{dr}{ds}^{\mu }\overline{A}_{\mu
}^{(pl)}(r),
\end{eqnarray}%
which represent the local contributions from the inertial, the external and
the plasma-collective EM field coupling terms, with $\overline{A}_{\mu
}^{(ext)}$ and $\overline{A}_{\mu }^{(pl)}$ denoting the corresponding
surface-averaged EM potentials (see Ref.\cite{EPJ2} for definition).
Furthermore, $L_{C}^{(RR)}$ represents the non-local (i.e., integral)
contribution arising from the EM self-field coupling, which is provided by%
\begin{equation}
L_{C}^{(RR)}(r,\left[ r\right] )=\frac{2q^{2}}{c}\frac{dr}{ds}^{\mu
}\int_{1}^{2}dr_{\mu }^{\prime }\delta (\widetilde{R}^{\mu }\widetilde{R}%
_{\mu }-\sigma ^{2}),  \label{LAGRANGIAN-SELF-EM}
\end{equation}%
where the $4-$scalar $\sigma ^{2}\equiv \xi ^{\mu }\xi _{\mu }$ is the
invariant radius of the surface distribution with respect to the center of
symmetry $r^{\mu }\left( s\right) $ having space-like radius $4-$vector $\xi
^{\mu }$, while $\widetilde{R}^{\mu }$ is a bi-vector defined as%
\begin{equation}
\widetilde{R}^{\alpha }\equiv r^{\alpha }\left( s\right) -r^{\alpha
}(s^{\prime }),
\end{equation}%
with $r^{\alpha }\left( s\right) $ and $r^{\alpha }(s^{\prime })$ denoting
the $4-$positions of the same center of symmetry at proper times $s$ and $%
s^{\prime }$ respectively.

Let us compute explicitly a canonical momentum $P$ conjugate to an arbitrary
coordinate $Q$ which depends only on the $4-$position. Here it is sufficient
to identify $Q$ with one of the components of $r^{\mu }$. Thus, the
generalized velocity $\frac{dQ}{ds}$ is simply one of the components of the $%
4-$velocity $u^{\mu }$. It follows that%
\begin{equation}
P=\frac{\partial L_{eff}}{\partial \left( \frac{dQ}{ds}\right) },
\end{equation}%
where contributions arise from all terms in Eq.(\ref{extremal Lagrangian-0}%
). Thus, $P$ must be identified with one of the components of the $4-$%
momentum $P^{\mu }$:%
\begin{equation}
P_{\mu }=m_{o}c\frac{dr_{\mu }(s)}{ds}+\frac{q}{c}\left[ \overline{A}_{\mu
}^{(ext)}+\overline{A}_{\mu }^{(pl)}+2\overline{A}_{\mu }^{(self)}\right] .
\label{pp}
\end{equation}%
Therefore, $P$ is conserved only if $Q$ is an ignorable coordinate for $%
L_{eff}$, namely if%
\begin{equation}
\frac{\partial L_{eff}}{\partial Q}=0.
\end{equation}%
If $Q=r^{k}$, with $k$ being one of the indices $0,1,2,3$, is ignorable by
assumption for $\overline{A}_{\mu }^{(ext)}(r)$ and $\overline{A}_{\mu
}^{(pl)}$, the only possible surviving contribution can arise from $%
L_{C}^{(RR)}$. In particular, it is found that%
\begin{equation}
\frac{\partial L_{C}^{(RR)}}{\partial Q}=\frac{\partial L_{C}^{(RR)}}{%
\partial r^{k}}=\frac{2q^{2}}{c}\frac{dr^{\mu }}{ds}\int_{1}^{2}dr_{\mu
}^{\prime }\left[ \frac{\partial }{\partial r^{k}}\delta (\widetilde{R}^{\mu
}\widetilde{R}_{\mu }-\sigma ^{2})\right] .  \label{dldq}
\end{equation}%
The integro-differential term in the previous equation can be shown to
generate a non-local coordinate-dependent contribution to $\frac{\partial
L_{C}^{(RR)}}{\partial Q}$ which ultimately depends on the bi-vector
component $\widetilde{R}_{k}$ . One finds%
\begin{equation}
\frac{\partial }{\partial r^{k}}\delta (\widetilde{R}^{\mu }\widetilde{R}%
_{\mu }-\sigma ^{2})=-\frac{\widetilde{R}_{k}}{\widetilde{R}^{\alpha
}u_{\alpha }(s^{\prime })}\frac{d}{ds^{\prime }}\left\{ \frac{\delta
(s-s^{\prime }-s_{ret})}{2\left\vert \widetilde{R}^{\alpha }u_{\alpha
}(s^{\prime })\right\vert }\right\} ,
\end{equation}%
where the retarded time $s_{ret}=s-s^{\prime }$ is the positive root of the
delay-time equation%
\begin{equation}
\widetilde{R}^{\mu }\widetilde{R}_{\mu }-\sigma ^{2}=0.
\end{equation}%
Substituting in Eq.(\ref{dldq}) then gives%
\begin{equation}
\frac{\partial L_{C}^{(RR)}}{\partial r^{k}}=\frac{q^{2}}{c}\frac{dr^{\mu }}{%
ds}\left\{ \frac{1}{\left\vert \widetilde{R}^{\alpha }u_{\alpha }(s^{\prime
})\right\vert }\frac{d}{ds^{\prime }}\left[ \frac{dr_{\mu }^{\prime }}{%
ds^{\prime }}\frac{\widetilde{R}_{k}}{\widetilde{R}^{\alpha }u_{\alpha
}(s^{\prime })}\right] \right\} _{s^{\prime }=s-s_{ret}}.  \label{symrr}
\end{equation}%
Let us investigate the general conditions under which the rhs of the
previous equation vanishes. This occurs manifestly if the quantity $\left\{ 
\frac{1}{\left\vert \widetilde{R}^{\alpha }u_{\alpha }(s^{\prime
})\right\vert }\frac{d}{ds^{\prime }}\left[ \frac{dr_{\mu }^{\prime }}{%
ds^{\prime }}\frac{\widetilde{R}_{k}}{\widetilde{R}^{\alpha }u_{\alpha
}(s^{\prime })}\right] \right\} _{s^{\prime }=s-s_{ret}}$ is orthogonal to
the instantaneous $4-$velocity $\frac{dr}{ds}^{\mu }$. This happens if the
first one is proportional to the instantaneous $4-$acceleration $\frac{d}{ds}%
u^{\mu }$. However, since $\widetilde{R}_{k}$ is only a component of $%
\widetilde{R}_{\mu }$, the previous orthogonality condition can at most hold
for a special subset of inertial reference frames. Therefore we conclude
that the condition of vanishing of $\frac{\partial L_{C}^{(RR)}}{\partial
r^{k}}$ cannot be cast in covariant way. Hence, even when there exists an
ignorable coordinate for the external and the plasma-collective fields, the
EM-RR term generally introduces explicit spatial dependences which do not
permit the existence of any form of exact symmetry, implying that exact
conservation laws can only be reached in absence of EM RR. Given validity of
the representation (\ref{extremal Lagrangian-0}) for the effective
Lagrangian $L_{eff}$ describing the dynamics of finite-size charged
particles, it follows that the existence of exact conservation laws is
generally prevented when the coupling term of the RR field $L_{C}^{(RR)}$ is
taken into account. The conclusion obtained here for finite-size charged
particles includes also the limiting case of point particles \cite{EPJ2}.

\end{document}